\documentclass[twocolumn,prd,amsmath,amssymb,showkeys,showpacs,superscriptaddress,floatfix,nofootinbib]{revtex4}
\usepackage{graphicx}

\begin{document}
\title{Searches for Cosmic-Ray Electron Anisotropies with the \textit{Fermi} Large Area Telescope}

\author{M.~Ackermann}
\affiliation{W. W. Hansen Experimental Physics Laboratory, Kavli Institute for Particle Astrophysics and Cosmology, Department of Physics and SLAC National Accelerator Laboratory, Stanford University, Stanford, CA 94305, USA}
\author{M.~Ajello}
\affiliation{W. W. Hansen Experimental Physics Laboratory, Kavli Institute for Particle Astrophysics and Cosmology, Department of Physics and SLAC National Accelerator Laboratory, Stanford University, Stanford, CA 94305, USA}
\author{W.~B.~Atwood}
\affiliation{Santa Cruz Institute for Particle Physics, Department of Physics and Department of Astronomy and Astrophysics, University of California at Santa Cruz, Santa Cruz, CA 95064, USA}
\author{L.~Baldini}
\affiliation{Istituto Nazionale di Fisica Nucleare, Sezione di Pisa, I-56127 Pisa, Italy}
\author{J.~Ballet}
\affiliation{Laboratoire AIM, CEA-IRFU/CNRS/Universit\'e Paris Diderot, Service d'Astrophysique, CEA Saclay, 91191 Gif sur Yvette, France}
\author{G.~Barbiellini}
\affiliation{Istituto Nazionale di Fisica Nucleare, Sezione di Trieste, I-34127 Trieste, Italy}
\affiliation{Dipartimento di Fisica, Universit\`a di Trieste, I-34127 Trieste, Italy}
\author{D.~Bastieri}
\affiliation{Istituto Nazionale di Fisica Nucleare, Sezione di Padova, I-35131 Padova, Italy}
\affiliation{Dipartimento di Fisica ``G. Galilei", Universit\`a di Padova, I-35131 Padova, Italy}
\author{K.~Bechtol}
\affiliation{W. W. Hansen Experimental Physics Laboratory, Kavli Institute for Particle Astrophysics and Cosmology, Department of Physics and SLAC National Accelerator Laboratory, Stanford University, Stanford, CA 94305, USA}
\author{R.~Bellazzini}
\affiliation{Istituto Nazionale di Fisica Nucleare, Sezione di Pisa, I-56127 Pisa, Italy}
\author{B.~Berenji}
\affiliation{W. W. Hansen Experimental Physics Laboratory, Kavli Institute for Particle Astrophysics and Cosmology, Department of Physics and SLAC National Accelerator Laboratory, Stanford University, Stanford, CA 94305, USA}
% ...

\collaboration{The \textit{Fermi} LAT Collaboration}
\noaffiliation

\begin{abstract}
The Large Area Telescope on board the \textit{Fermi} satellite (\textit{Fermi} LAT) detected more 
than 1.6 $\times$ $10^6$ 
cosmic-ray electrons/positrons with energies above 60~GeV during its first year of operation. 
The arrival directions of these events were searched for anisotropies 
of angular scale extending from $\sim$10$^\circ$ up to 90$^\circ$, and of 
minimum energy extending from 60~GeV up to 480~GeV. Two independent techniques were
used to search for anisotropies, both resulting in null results. Upper limits on the degree of the anisotropy were set that
depended on the analyzed energy range and on the anisotropy's angular scale. The upper limits for a dipole anisotropy ranged from $\sim0.5\%$ to $\sim10\%$.
\end{abstract}



\maketitle

\section{Introduction}

The majority of detected high-energy (GeV--TeV) charged primary Cosmic Rays (CRs) 
is believed to be produced in our galaxy, most likely in Supernova Remnants (SNRs).
During the transport from their source of origin to our solar system, CRs scatter on random
and irregular components of the $\mu$G Galactic Magnetic Field (GMF), which almost isotropize
the CRs' direction distribution. This happens because the Larmor radius for a typical value of 4$\mu$G for the GMF
and for a 100~GeV singly-charged particle
is $\sim3\times10^{-5}$~pc, considerably smaller than the typical distance to a nearby source (of the order of a hundred pc).
Nevertheless, several ground experiments across a wide
range of energies have still detected anisotropies of medium ($\sim10^\circ$) to large ($90^\circ$) angular scales
(for instance see \cite{milagro_smallscale,milagro_largescale,tibet06,icecube,argo,superK}). Searches for such anisotropies
can provide unique information on the sources of CRs and the environment in which they have
propagated. 

Contrary to hadronic CRs, high-energy ($>$GeV) Cosmic Ray Electrons and
Positrons (CREs) propagating in the GMF lose their energy rapidly through
synchrotron radiation and by inverse Compton
collisions with low-energy photons of the interstellar radiation field.
CREs observed with energies 100~GeV (1~TeV) originated from relatively nearby
locations, less than about 1.6~kpc (0.75~kpc) away \cite{foot_1}.
This means that it could be possible that such high-energy CREs originate from
a highly anisotropic collection of a few nearby sources (possibly
pulsars and SNRs). 
Therefore, depending on the propagation properties in the GMF, the detection of excess CREs with energies high
enough to minimize both the geomagnetic field and any heliospheric
effects might reveal the presence of such nearby CRE sources. Similarly, assuming a reasonable
distribution of nearby CRE sources, measurements of the CRE anisotropy can be used
to constrain the diffusion of CREs in the galaxy.
Finally, further anisotropy that is not associated with nearby CR sources is expected
to result from the Compton-Getting (CG) effect \cite{CG1935}, in which 
the relative motion of the observer with respect to the CR plasma changes the 
intensity of the CR fluxes, with larger intensity arriving from the direction of
motion and lower intensity arriving from the opposite direction. 

The dataset of the Large Area Telescope (LAT) on board the \textit{Fermi}
satellite \citep{electronpaper} provides a unique high statistics sample for CRE
anisotropy studies. In this paper, we report the 
results of a search for anisotropies in the reconstructed directions of the events
in this dataset.

\section{The Instrument and The Data}

The LAT is a pair-conversion gamma-ray telescope designed to measure gamma rays in the energy range
from 20~MeV to more than 300~GeV. In this paper a brief description of
the LAT is given, while full details can be found in \cite{Atwood2009}.

The LAT is composed of a $4 \times 4$ array of 16 identical towers designed to
convert incident gamma-rays into $e^{+} e^{-}$ pairs, and to determine their
arrival directions and energies. Each tower hosts a tracker module and a
calorimeter module. Each tracker module consists of 18 x-y planes of
silicon-strip detectors, interleaved with tungsten converter foils, for a
total on-axis thickness equivalent to 1.5 radiation lengths (r.l.). Each
calorimeter module, 8.6 r.l. on-axis thick, hosts 96 CsI(Tl) crystals,
hodoscopically arranged in 8 perpendicular layers. The instrument is
surrounded by a segmented anti-coincidence detector that tags the
majority of the charged-particle background. 

Although the LAT was designed for detecting photons, it was recognized that it can
also work as an excellent detector of high-energy CREs. The calorimeter
behaves in the same way for electrons and gamma rays, and helps with
discriminating electromagnetic from hadronic showers. The early part of
the shower can be reconstructed in detail in the tracker, contributing
to the hadron/electron separation. 

The analyzed dataset corresponds to the first year of LAT science operation and 
starts on August 2008. Because the Earth's magnetic field can affect the directions of incoming
CRE events, introducing or hiding anisotropies, we have selected events with an energy high enough (E$>$60~GeV)
to minimize the geomagnetic field's influence yielding $\sim1.6$~million events. 
While this 60~GeV energy threshold is significantly higher than the geomagnetic cut-off in any part of \textit{Fermi's} orbit, 
the geomagnetic field could still introduce some deflections even above this energy, possibly giving rise to a spillover effect in adjacent regions of the sky.
Furthermore, the Heliospheric Magnetic Field (HMF) can also affect the directions of particles propagating in it, exhibiting
an appreciable effect on CREs with energies $\lesssim100$~GeV, which diminishes quickly with rising energy, until it 
becomes negligible at energies over several hundred GeV. However, it is not easy to quantify the HMF's influence on CREs 
of some energy propagating in it, since this would require a good knowledge of the HMF's structure and of the propagation of CREs through it. 
Because of the absence of a well defined energy threshold over which the HMF effects can be ignored, we decided to
start the analysis from 60~GeV with the caveat that some of our lower-energy results might be affected by the HMF.

To minimize the contamination from the Earth's albedo, events with reconstructed
directions near the Earth's limb or detected during the interval when the Earth's limb enters 
the LAT's field of view more than usual (when the rocking angle is greater than $43^\circ$) were 
rejected. This cut removed roughly $16\%$ of the events, leaving a total of $\sim1.35$ million
events for this analysis. It should be noted that all the events in the dataset
were detected while the \textit{Fermi} spacecraft was outside the South Atlantic Anomaly.

The \textit{Fermi} satellite is usually operated in ``sky-survey mode''. In this
operation mode, the sky is fully observed every 2~orbits ($\sim$3 hours). This
ensures a uniform (up to $\sim15\%$) all-sky exposure and allows us 
to search for anisotropies of any angular scale (including dipole anisotropy) 
and from any direction in the sky.

The angular resolution of the LAT for E$>$60~GeV electron events is about $0.1^\circ$ or better and
the energy resolution (1$\sigma$) is $\sim10\%$. The contamination of the analyzed dataset with
other species, such as photons or protons, can result in some systematic uncertainties. The fraction
of hadron events in the dataset ranges from $\sim4\%$ at 20~GeV to $\sim20\%$ at 1~TeV, while
the photon contamination is negligible ($<0.1\%$). The full details of \textit{Fermi's} CRE data analysis can be found in \cite{fullelectronpaper}.

\section {Method}

The whole sky was searched for anisotropies in Galactic coordinates with no a priori assumptions
on the direction, the angular scale, or the energy spectrum of a potential signal. The dataset was analyzed in its entirety,
without trying to divide it in time to perform a search optimized for detecting transient anisotropies.

One way to search for anisotropies is to first calculate the flux of CRE particles 
from each direction in the sky (equal to the ratio of the number of detected events from some direction
over the exposure towards the same direction), and then examine its directional distribution.
The flux calculation, which requires knowledge of the exposure, depends on the effective area of the detector 
and the accumulated observation livetime. 
The effective area, calculated from a Monte Carlo simulation of the 
instrument, could suffer from systematic errors, such as a dependence
on the time or on the location of the spacecraft, 
or any miscalculations of the dependence of the effective area on the instrument coordinates (off-axis 
and azimuthal angle). Naturally, any systematic errors involved in the calculation of the exposure 
will propagate to the flux, possibly affecting its directional distribution. 
If the magnitude of these systematic errors is comparable to or larger than
the statistical power of the available dataset, 
their effects on the flux's directional distribution might masquerade as a real detectable anisotropy. 
As will be shown below, the statistics of the available dataset allow us 
to search for anisotropies as small as a fraction of a percent. 
Because the effective area of the detector is not known to a better accuracy, it follows that it is not safe to perform
anisotropy searches of such sensitivity by examining the directional distribution of
the particle flux. For that reason, we employed two alternative methods that are free of systematic errors
larger than a fraction of a percent. 

\subsection{No-anisotropy map creation}
The starting point of this analysis
is the construction of a sky map that shows how the sky as seen by the \textit{Fermi}-LAT would look
on average if the CRE direction distribution were perfectly isotropic. 
This sky map, hereafter called ``no-anisotropy sky map'', represents 
the null hypothesis for the existence of an anisotropy. A comparison of the no-anisotropy sky map 
to a sky map generated by the actually detected CRE events (the ``actual'' or ``signal'' sky map)
was used to reveal the presence of any anisotropies in the data. 

As a cross-check of the systematic errors involved in this analysis,
two techniques were used to build the no-anisotropy sky map, producing similar results.
The two techniques are described below.

\begin{itemize}
\item \textbf{Shuffling technique}:\\
One way of generating the no-anisotropy map is to randomize the reconstructed directions of the detected
events \citep{Cassiday2009, Bird1999, Ambrosio2003}. In case the direction distribution of the CR flux is perfectly isotropic, a time-independent intensity should be detected when looking at any given detector direction. Possible time variation of the intensity would be due only to changes in the operating conditions of the instrument. A set of isotropic simulated events can
be built by randomly coupling the times and the directions of real
events in local instrument coordinates. The randomization is performed starting with the position of a given event in the LAT frame and exchanging it with the direction of another event, which was selected randomly from the data set with a uniform probability. Starting with this information, the sky direction is re-evaluated for the simulated event. In this way the random coupling preserves the exposure and the total number of events.
This process is repeated
multiple times (100), with each time producing a sky map that is compatible with an isotropic CRE
direction distribution. The final no-anisotropy sky map is produced by taking the average of these sky maps.
By this construction, the simulated data set preserves exactly the
energy and angular (with respect to the LAT reference frame) distributions, and also
accounts for the detector dead times.
To minimize the possible effects of a varying CRE event rate, the data set is first split
into 10 segments with equal number of events, and the technique is applied to each of these sets
separately.

\item \textbf{Direct-integration technique}: \\
This technique is based on \citep{Atkins2003}. In general, the rate of events detected
in some narrow solid angle around a given direction ($\theta$, $\phi$) \cite{foot_2} at some
time $t$ is equal to the all-sky rate at that time $R_{allsky}(t)$ times
the probability $P(\theta,\phi,t)$ of an event being reconstructed inside that same
solid angle. Given a dataset and the LAT's pointing information, we 
can calculate the values of the two functions $R_{allsky}(t)$ and $P(\theta,\phi,t)$.
Similarly, given the values of these two functions and the LAT's pointing information for some observation
we can predict how the sky would look for that same observation (i.e. construct a sky map).
The main idea of this method is to first extract from the CRE dataset the set of values 
of $R_{allsky}(t)$ and $P(\theta,\phi,t)$ that corresponds to an isotropic CRE distribution,
and then construct the associated no-CRE-anisotropy sky map. The presence of any anisotropies
in the data would create transient fluctuations in the instantaneous values of these functions,
as these anisotropies passed through the LAT's field of view. However, these anisotropies would
have no effect on the longer-term average values of these functions, since any
transient fluctuations would be averaged out. In this application,
we used a constant-over-time $P_{ave}(\theta, \phi)$ produced after averaging $P(\theta,\phi,t)$ over the whole dataset.
For an isotropic sky, the only time dependence of $P(\theta,\phi,t)$ would come only from temporal variations of 
the dependence of the detector's effective area on $\theta$ and $\phi$. We did not detect any such
variations in the data, therefore using an average over the whole dataset $P_{ave}(\theta, \phi)$ was valid.
Similarly, any temporal variations of the effective area of the detector would create fluctuations of
the instantaneous value and the longer-term averages of $R_{allsky}(t)$, depending on the time
scales of the effective-area variations. Unlike $P(\theta,\phi,t)$, which remained sufficiently constant for this purpose, 
the all-sky rate exhibited fluctuations on multiple time scales caused by varying background rates occurring as the
spacecraft was moving through regions of different geomagnetic coordinates and by changes in the instrument's hardware settings.
Any such instrumental effects affecting the all-sky rate were parametrized, and given an averaged-over-multiple-orbits value of the all-sky rate \cite{foot_3},
the all-sky rate at some shorter-duration segment could be accurately predicted. These parametrizations are necessary
because the direct-integration method constructs the no-anisotropy-skymap incrementally, adding the results from 
observations of short enough duration that the LAT's pointing can be assumed quasi-constant (e.g., 30~s long).
As a cross-check, this technique was also applied with $R_{allsky}(t)$
being given by the instantaneous rate of actually detected CRE events, instead of an averaged value. This choice has the benefit of automatically 
taking care of any temporal variations of the effective area, avoiding the need to apply any corrections and, most importantly, 
avoiding any systematic errors introduced by such corrections. It should be noted however that not performing
an averaging in the all-sky rate weakens the power of this method to smear out the presence of 
any anisotropies in the data, with the result of any stronger anisotropies possibly leaking in the 
no-CRE-anisotropy sky map. Nevertheless, for this application, the results of the direct-integration technique when
using the instantaneous event rate and when using the averaged-over-multiple
orbits event rate plus the necessary corrections were consistent with each other. 
\end{itemize}

For our sky maps, we adopted the HEALPix \cite{healpix} pixelization scheme to ensure that all pixels
across the sky have the same area (or solid angle).
The resolution of the HEALPix grid is expressed by the parameter $N_{side}$, which defines
the total number of pixels: $N_{pix} = 12 \times N_{side}^2$. 
Our maps had $N_{side}=32$, which corresponds to 12,288 $\sim3\,$deg$^{2}$ pixels. 
The sky maps presented in this paper are all in Galactic coordinates. 
The no-anisotropy maps were compared to the actual sky map using two independent methods, described below.

\subsection{Direct bin-to-bin comparison}
The first method is a simple direct bin-to-bin comparison of the two maps, in which a search for 
statistically significant deviations between the number of actually
detected and the number of expected events under the assumption of isotropy is performed. 

One way to search for anisotropies of some angular scale, is to use sky maps composed of \textit{independent} bins
with bin size similar to the angular scale of the anisotropy under search.
However, when using independent bins, a potential anisotropic signal might become too weak to
be detected since it will probably be distributed among multiple adjacent bins. 
A more sensitive way to perform the search, adopted in the present analysis, is to use sky maps consisting of a large number of \textit{correlated} bins.
The content of a correlated bin is equal to the integrated number of events in a
circular region around that bin. Using such ``integrated sky maps'', it is very likely that there will be
at least one bin with its center roughly aligned with the direction of the center of a potential anisotropy, reducing spillover
effects and increasing sensitivity. 
In general, the sensitivity for detecting an anisotropy of given angular scale is greater
when an integration radius close to that scale is chosen. 
If the integration radius is too small or too large compared to the angular scale of the prospective anisotropy,
the sensitivity becomes sub-optimal since either the signal can be split among
several adjacent bins or there can be too much ``background'' (isotropic signal) contamination. 
In this work, to search for anisotropies of various angular scales we compared multiple pairs of integrated no-anisotropy and actual sky maps,
with each pair corresponding to a different integration radius. The integration radii were chosen to cover the range of 
anisotropy angular-scales under consideration and ranged from $10^\circ$ to $90^\circ$.

The comparison was performed by calculating the statistical significance of the difference between
the contents of the integrated actual and the no-anisotropy sky maps. 
To take into account the statistical errors involved in construction of the no-anisotropy sky maps 
by the event-shuffling technique, the significances for that method were evaluated using  
the prescription in ~\cite{LiMa}, by using the likelihood method \cite{foot_4}. It should be noted that the 
results of the event-shuffling technique were stable even after a couple tens 
of repetitions. For this application, we used 100 repetitions, rendering the associated statistical errors
negligible. For the case of the direct-integration technique, since its produced no-anisotropy sky map is the
result of a direct calculation, there were no associated statistical errors and the significance
was calculated using simple Poisson probabilities.
 
\begin{figure}[ht]
\includegraphics[width=1\columnwidth,keepaspectratio,clip,trim=0 0 15 0]{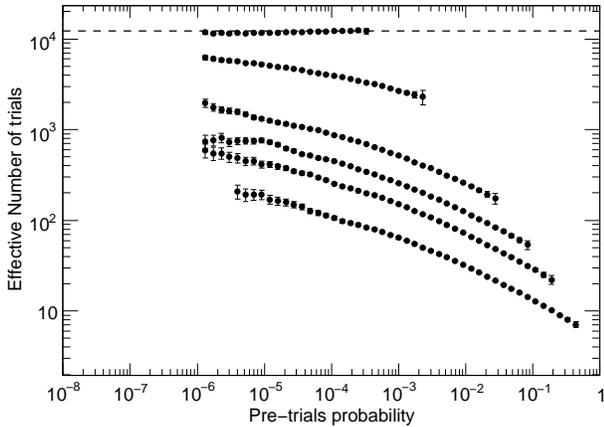}
\caption{Curves: Number of effective trials ($T_{eff}$) involved in evaluating the contents of a single significance map. From top to bottom: trials for an independent-bins significance map, and trials for integrated maps of 10$^\circ$, 30$^\circ$, 45$^\circ$, 60$^\circ$, and 90$^\circ$ integration radius. The horizontal dashed line shows the number of bins in these maps (12,288). As expected, the number of effective trials for an independent-bins sky map (top curve) is equal to the number of bins in the map (horizontal line). The effective numbers of trials for integrated maps are in general smaller than the number of bins, with the integration radius being inversely correlated to the number of effective trials. The error bars show the statistical error that arises from the finite number of simulated sky maps.}
\label{fig:Trials}
\end{figure}

As was mentioned above, this is a search for anisotropies from any direction in the whole sky. As such, 
it involves a large number of trials (independent tests), that have to be accounted for when judging the statistical
significance of its results. For a whole-sky search performed using independent-bin sky maps,
the number of trials is equal to the number of pixels in the sky map. On the other hand, 
searches that use correlated-bin sky maps, such as this one, involve a number of \textit{effective trials} that is in general
smaller than the number of bins in one such sky map. The larger the integration radius, the higher the degree of 
correlation between adjacent bins and the smaller the number of effective trials per evaluated bin.

The number of effective trials involved in evaluating the contents of an integrated map was evaluated
using a Monte Carlo simulation. Fake significance maps were built
corresponding to various integration radii and to a perfectly isotropic signal. By counting the fraction
of such sky maps $P_{post}$ that contained at least one bin with a probability smaller than some
value $P_{pre}$, we calculated the effective number of trials involved in the search of sky maps of some integration radius
\cite{foot_5}:

\begin{equation}
T_{eff}=\frac{log(1-P_{post})}{log(1-P_{pre})}.
\label{eq:teff}
\end{equation}

The effective number of trials for each integration radius and for the case of an independent-bins sky map is shown in Fig.~\ref{fig:Trials}. Using this effective number of trials, the post-trials probability $P_{post}$ corresponding to a pre-trials $P_{pre}$ probability can be calculated as:
\begin{equation}
P_{post}=1-\left(1-P_{pre}\right)^{T_{eff}}. 
\label{eq:p_post}
\end{equation}

\subsection{Spherical harmonic analysis}
A more robust method involves a spherical harmonic analysis of a ``fluctuations sky map'' equal to the ratio of the
actual and no-anisotropy sky maps minus one. Initially, the fluctuations sky map is expanded in the basis of
spherical harmonics, producing a set of coefficients $a_{lm}$. Then, an angular
power spectrum is constructed by calculating the average variance of the $2l+1$ $a_{lm}$ coefficients at each multipole $l$ as

\begin{equation}
\hat{C}_l=\frac{1}{2l+1}\overset{l}{\underset{m=-l}{\text{\ensuremath{\sum}}}}\left|a_{lm}\right|^2.
\end{equation}

The power spectrum characterizes the intensity fluctuations
as a function of the angular scale. An increased power $\hat{C}_{l}$ at a multipole $l$ corresponds to
an anisotropic excess of angular scale $\sim180^{\circ}/l$.
The spherical harmonic analysis was performed using the \textit{anafast} code provided with  
the HEALPix tools \cite{healpix}.

The data can be treated as the sum of two independent components: an anisotropic component, which we are trying to detect, and an isotropic component, which 
is known and equivalent to white noise. Because these two components are not correlated, 
the \textit{observed} angular power spectrum can be similarly split into two components: $\hat{C}_{l}=\hat{C}_{l}^{aniso}+\hat{C}_{l}^{N}$.
The quantities $\hat{C}_{l}$, $\hat{C}_{l}^{aniso}$ and $\hat{C}_{l}^{N}$ are random variables, in general different than their 
true underlying quantities $C_{l}$, $C_{l}^{aniso}$, and $C_{l}^{N}$ respectively. Specifically,
the observed quantities follow a $\chi^{2}_{2l+1}$ distribution centered at their corresponding true values. 
The true value of the isotropic (white noise) component is \cite{cuoco}:

\begin{equation}
C_{l}^{N}=\frac{4\pi}{N},
\label{eq:ClExp}
\end{equation}
where $N$ is the total number of observed events.

To search for anisotropies in the data, we examined the validity of the null hypothesis: $C_{l}^{aniso}=0$ or equivalently $C_{l}=C_{l}^{N}$.
This was accomplished by checking whether the observed power spectrum $\hat{C}_{l}$ was statistically compatible with the known true value of the isotropic power spectrum $C_{l}^{N}$. 

The resulting power spectra can also be used for setting upper limits on the degree of anisotropy:

\begin{equation}
 \delta\equiv\frac{I_{max}-I_{min}}{I_{max}+I_{min}},
\end{equation}
where $I_{max}$ and $I_{min}$ are the maximum and minimum values of the CRE intensity. Consider a dataset
consisting of the sum of a perfectly isotropic signal of constant intensity $I_{0}$ and of a dipole anisotropy of maximum intensity
$I_{1}$. The overall intensity at an angular distance $\theta$ from the maximum of the dipole anisotropy will be $I(\theta)=I_{0}+I_{1}cos(\theta)$.
For this dataset, the degree of its dipole anisotropy is 

\begin{equation}
\delta=\frac{I_{1}}{I_{0}}.
\label{eq:dip}
\end{equation}
The fluctuation map describing this dataset is:

\begin{equation}
 f(\theta)=\frac{I(\theta)-<I(\theta)>}{<I(\theta)>}=\frac{I(\theta)-I_{0}}{I_{0}}=\frac{I_1}{I_0}cos(\theta).
\label{eq:ftheta}
\end{equation}
Since $Y^{0}_{1}(\theta,\phi)=\sqrt{\frac{3}{4\pi}}cos(\theta)$, it follows from Eq.~\ref{eq:ftheta}
that
$f(\theta)=\left( \frac{I_1}{I_0} \sqrt{\frac{4\pi}{3}} ~ \right) \times Y^{0}_{1}$
or that 

\begin{figure}[ht!]
\includegraphics[width=0.95\columnwidth,keepaspectratio,clip]{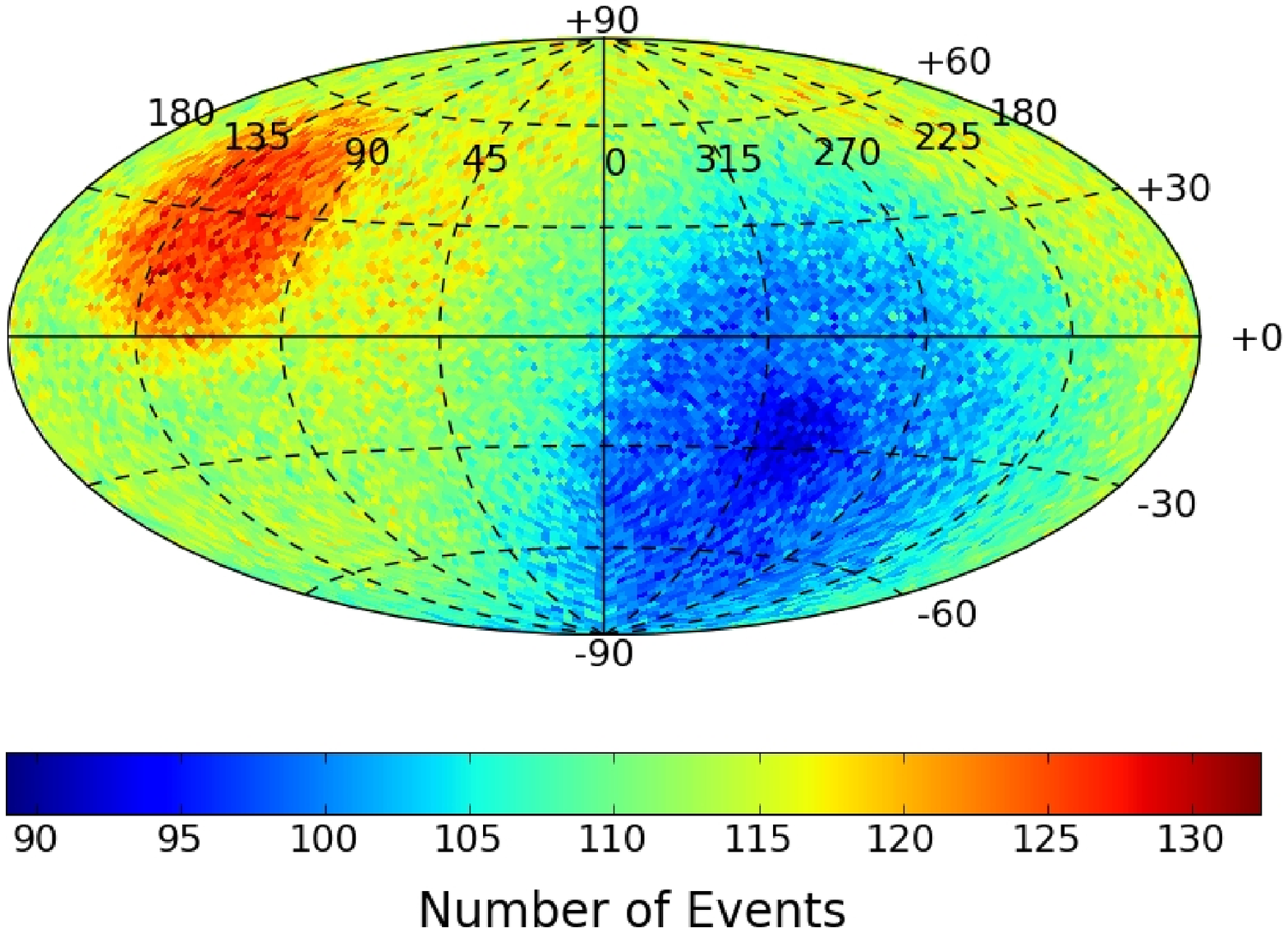}
\includegraphics[width=0.95\columnwidth,keepaspectratio,clip]{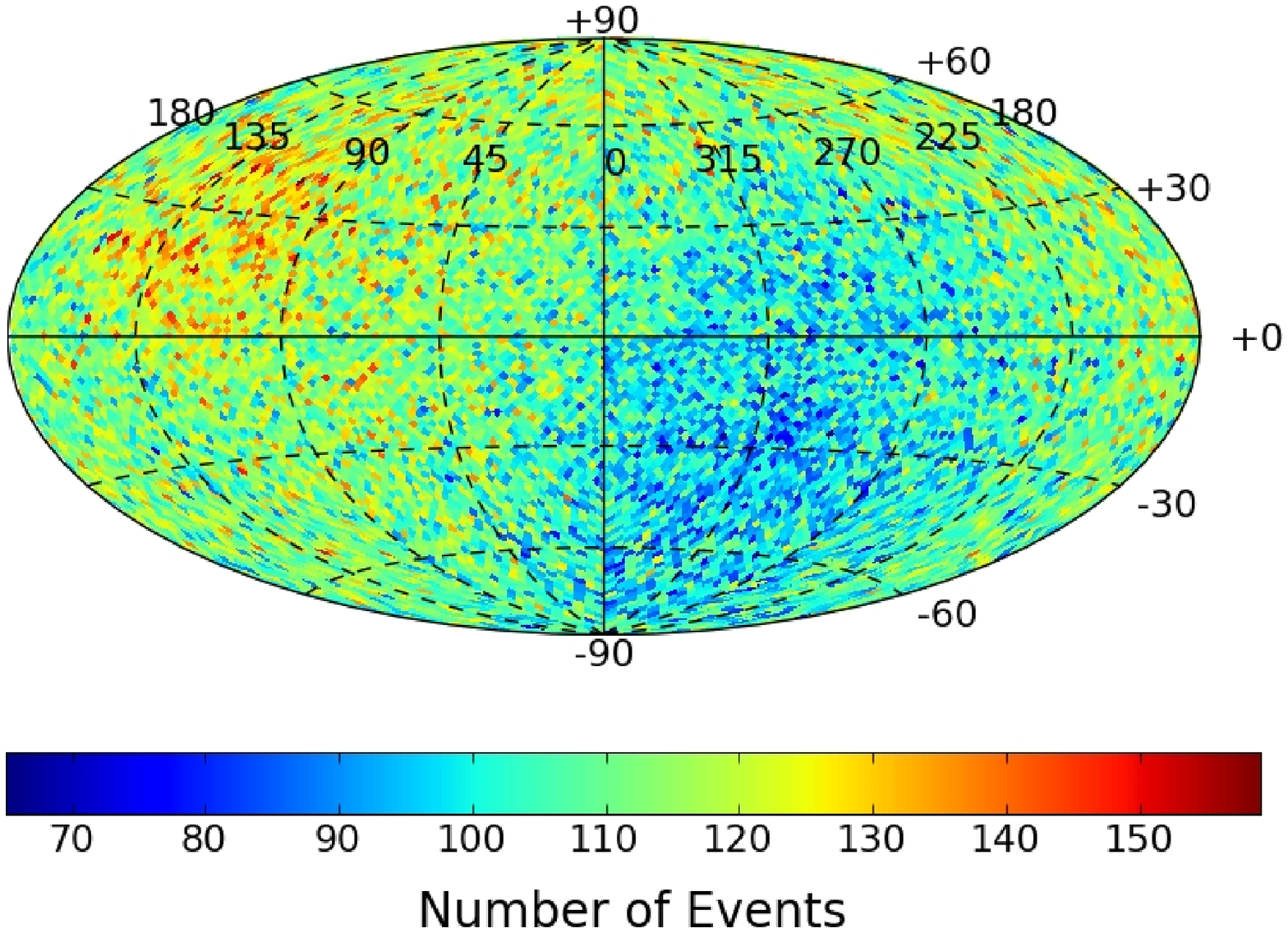}
\includegraphics[width=0.95\columnwidth,keepaspectratio,clip]{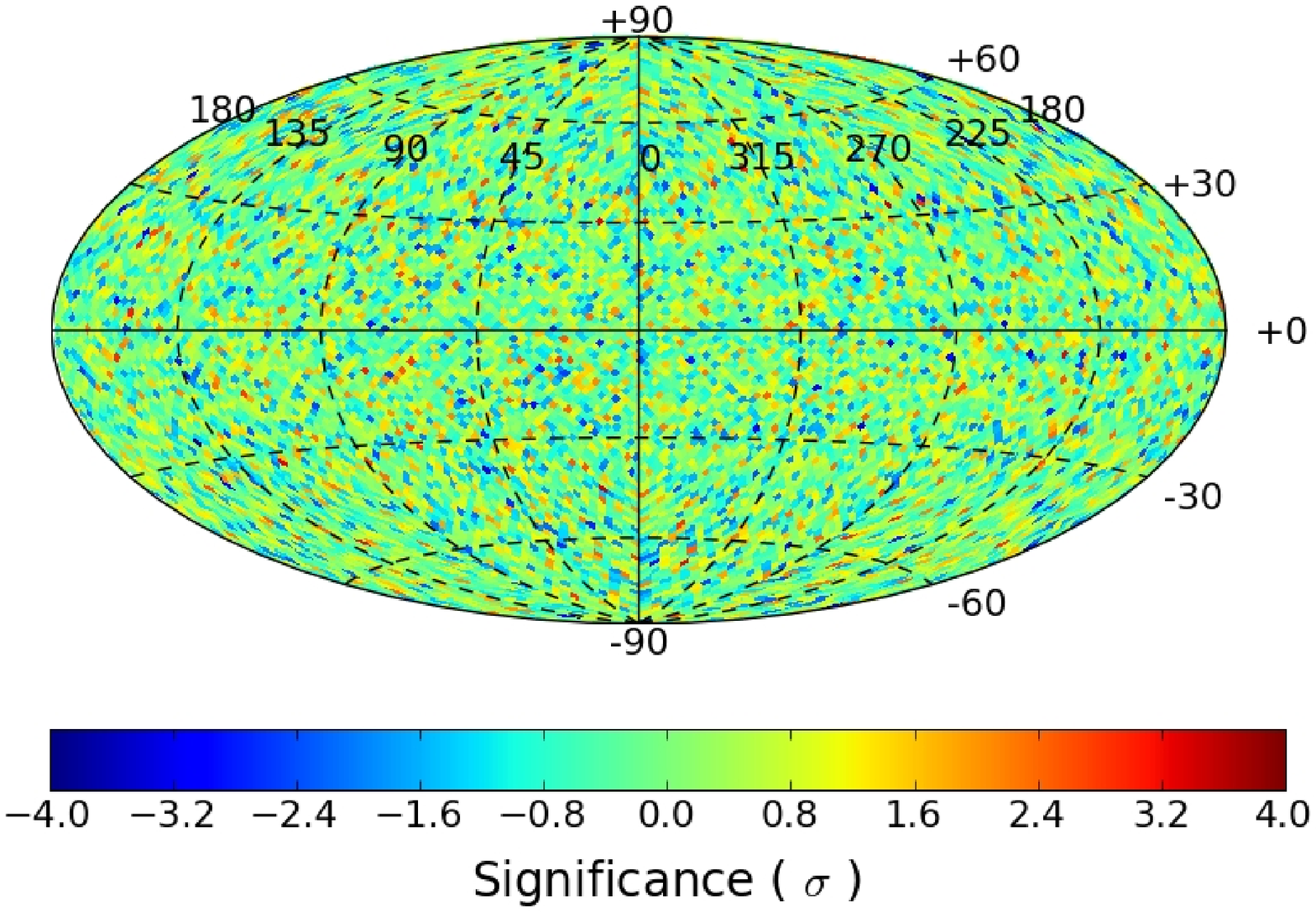}
\caption{From top to bottom: no-anisotropy sky map for E$>$60~GeV; sky map of actually detected CREs with E$>$60~GeV; significance map produced by comparing the above maps. The shape of the actual and no-anisotropy sky maps results from the fact that the sky was not observed with uniform exposure.}
\label{Fig_independent_60GeV}
\end{figure}

\begin{equation}
a_{10}=\frac{I_1}{I_0}\sqrt{\frac{4\pi}{3}}.
\end{equation}
All but the $l=1$ term of the power spectrum vanish. The non-zero term is:

\begin{equation}
C_1\equiv\frac{1}{3}\sum_{m=-1}^{1}|a_{1m}|^2=\frac{a_{10}^2}{3}=\left(\frac{I_1}{I_0}\right)^2\frac{4\pi}{9}.
\label{eq:C1}
\end{equation}
It should be noted that since $C_1$ (as all the $C_l$) is a rotationally invariant quantity, the above expression for $C_1$, derived for the reference frame in which $a_{11}=0$ and $a_{1-1}=0$, is valid in general for every dipole and reference frame.
From Eq.s \ref{eq:dip} and \ref{eq:C1} we derive a relation between the degree of dipole anisotropy and the value of the dipole power:

\begin{equation}
 \delta=3\sqrt{\frac{C_1}{4\pi}}.
\end{equation}

To set an upper limit on $\delta$ we start by calculating the probability distribution function (pdf) of $\hat{\delta}=3\sqrt{\frac{\hat{C}_1}{4\pi}}$ by a change of variable on the probability distribution function of $\hat{C}_{1}$ ($\chi^{2}_{3}$ centered on $C_{1}$). The probability of observing
a certain $\hat{C}_{1}$ given the true dipole power $C_{1}$ (i.e. the
probability density function to observe $\hat{C}_{1}$ given the true
dipole power $C_{1}$) is: 

\begin{equation}
 P(\hat{C}_{1};C_{1})=\frac{3\sqrt{3}}{\sqrt{2\pi}C_1}\sqrt{\frac{\hat{C_{1}}}{C_1}}exp\left(-\frac{3\hat{C}_1}{2C_{1}}\right).
\end{equation}
With a change of variables we obtain:

\begin{equation}
 P(\hat{\delta};\delta)=\frac{3\sqrt{6}}{\sqrt{\pi}}\frac{\hat{\delta}^2}{\delta^3}exp\left(-\frac{3\hat{\delta}^2}{2\delta^2}\right).
\end{equation}

The upper limit of the true anisotropy $\delta_{UL}$ is the value of $\delta$ for which the integrated probability of measuring a value of $\hat{\delta}$ at least as large as the one we measured is equal to the confidence level. Specifically, for a confidence level CL, the upper limit on the dipole anisotropy can be evaluated using the frequentist approach \cite{pdg} by solving:

\begin{equation}
\int_{0}^{\hat{\delta}_{\textrm{meas}}}P(\hat{\delta};\delta) d \hat{\delta} = 1-CL,
\end{equation}
where
$\hat{\delta}_{\textrm{meas}}\equiv3\sqrt{\frac{\hat{C}_{1,\textrm{meas}}}{4\pi}}$
is the dipole anisotropy that corresponds to our single measurement of the
dipole power spectrum $\hat{C}_{1,\textrm{meas}}$. Integrating and solving for $\delta$ we obtain $\delta_{UL}$.

\begin{figure}[ht!]
\includegraphics[width=1.0\columnwidth,keepaspectratio,clip]{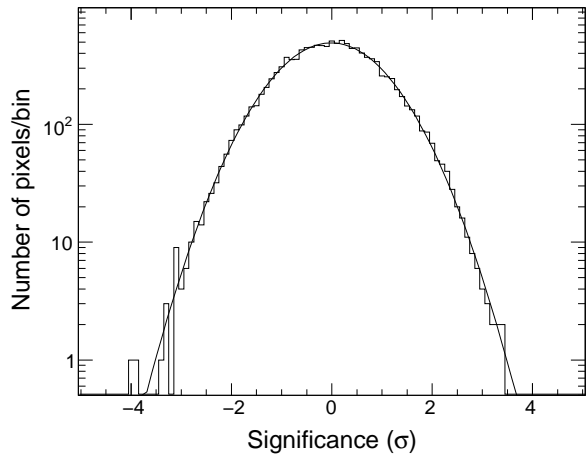}
\caption{Distribution of the significance values of the sky map in the third panel of Fig. \ref{Fig_independent_60GeV} (histogram) 
together with a best fit with a Gaussian function (solid line).}
\label{fig3}
\end{figure}

\section{Results}
In this section we report the results of our search for a steady excess
of CREs from any direction of the sky. 

The two techniques, event shuffling and direct integration, produced similar no-anisotropy sky maps,
and any differences were considerably smaller than the smallest amplitude of a detectable signal.
For brevity and unless otherwise noted, all the sky maps and upper limits presented
in this paper were produced using the no-anisotropy sky map of the event shuffling technique. 

\begin{figure*}[h!]
\includegraphics[width=0.87\columnwidth,keepaspectratio,clip]{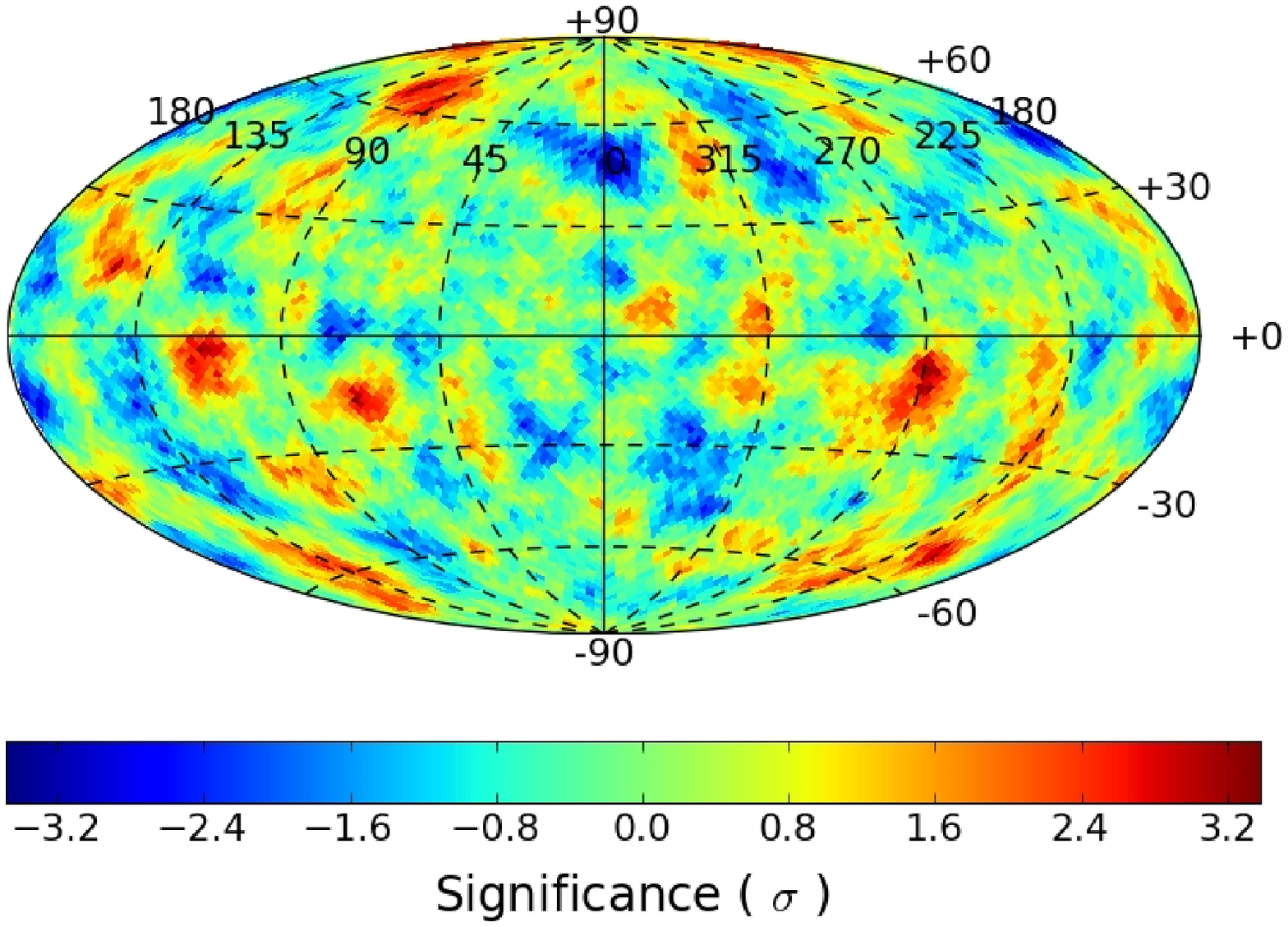}
\includegraphics[width=0.87\columnwidth,keepaspectratio,clip]{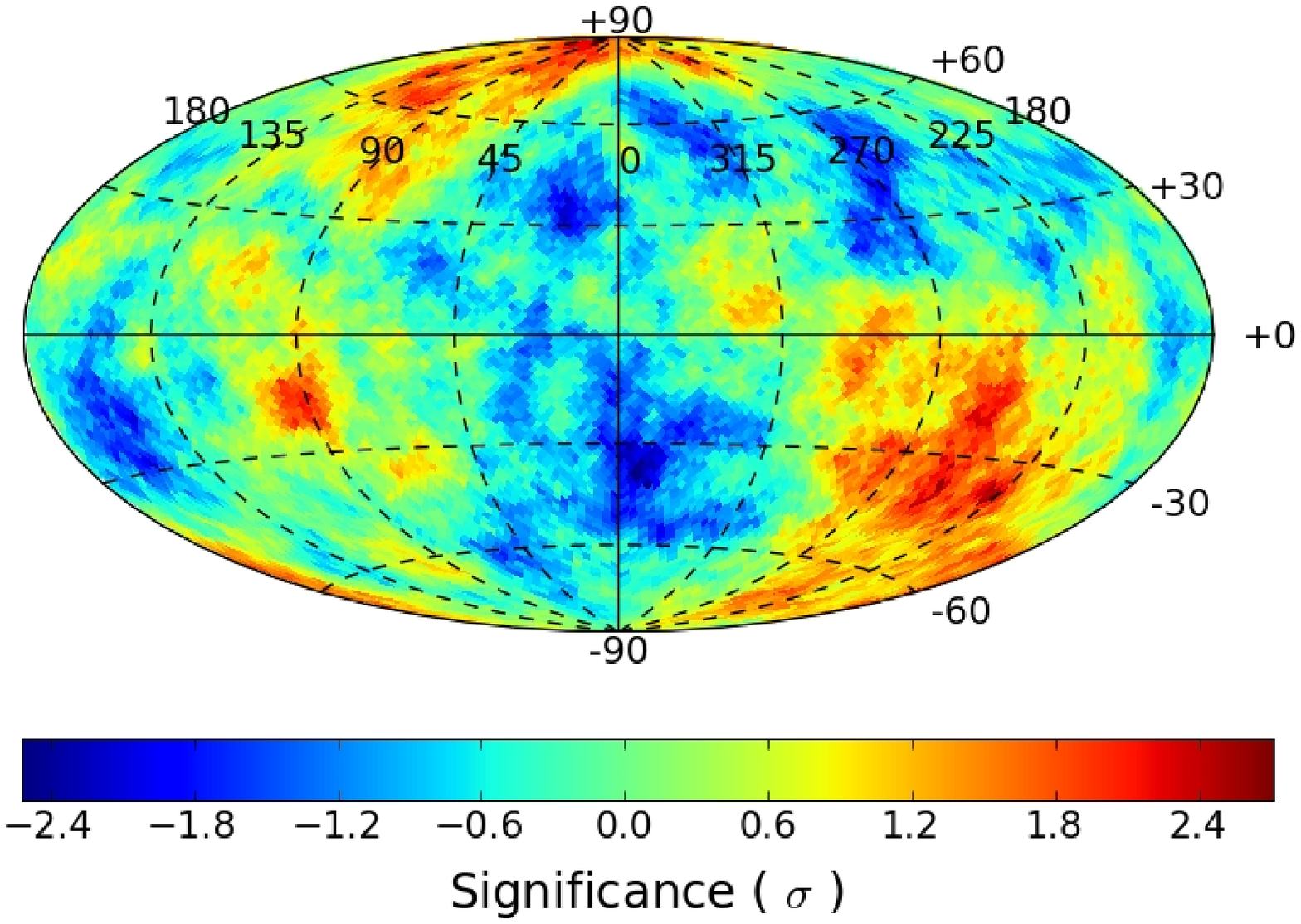}
\includegraphics[width=0.87\columnwidth,keepaspectratio,clip]{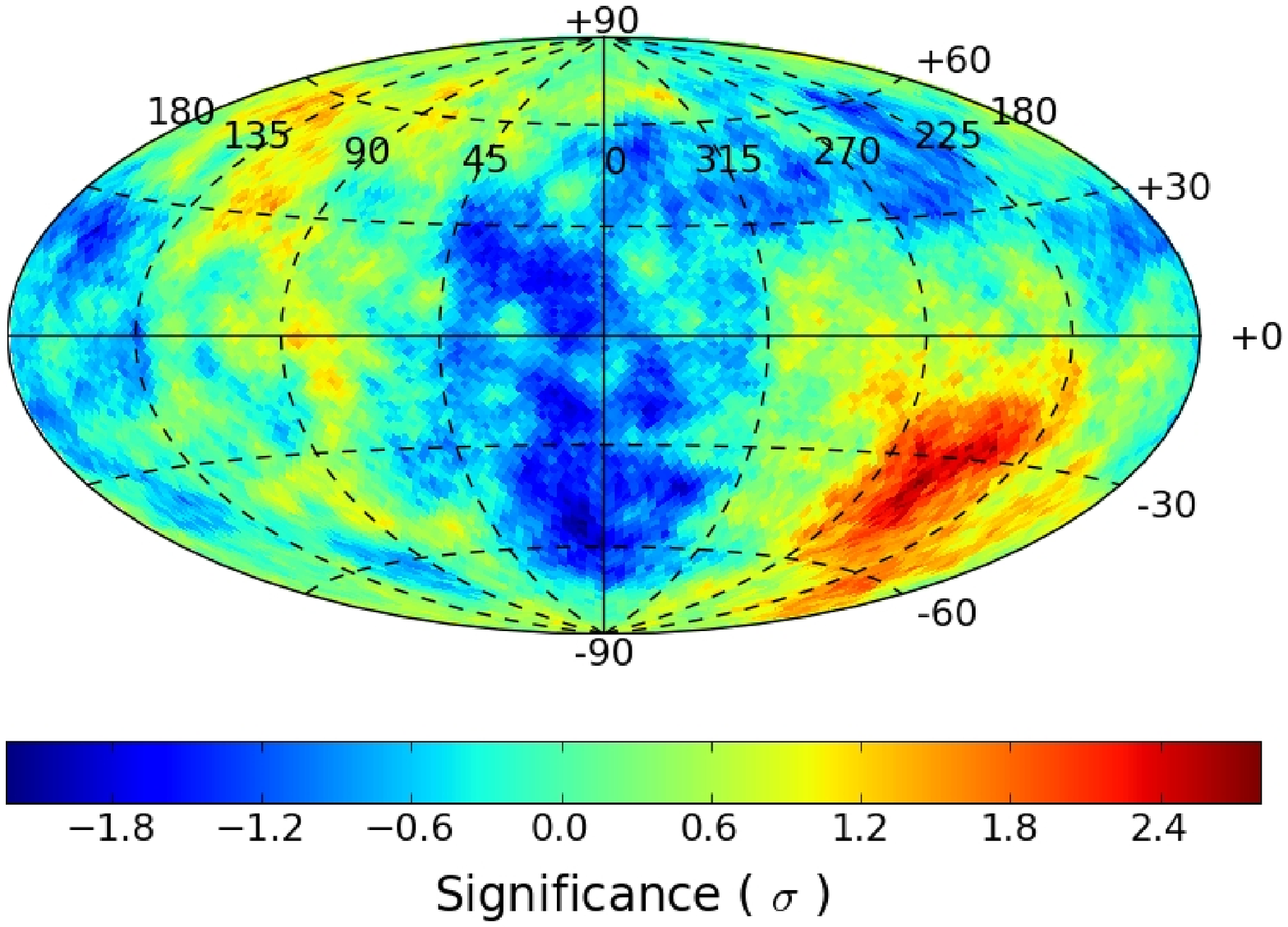}
\includegraphics[width=0.87\columnwidth,keepaspectratio,clip]{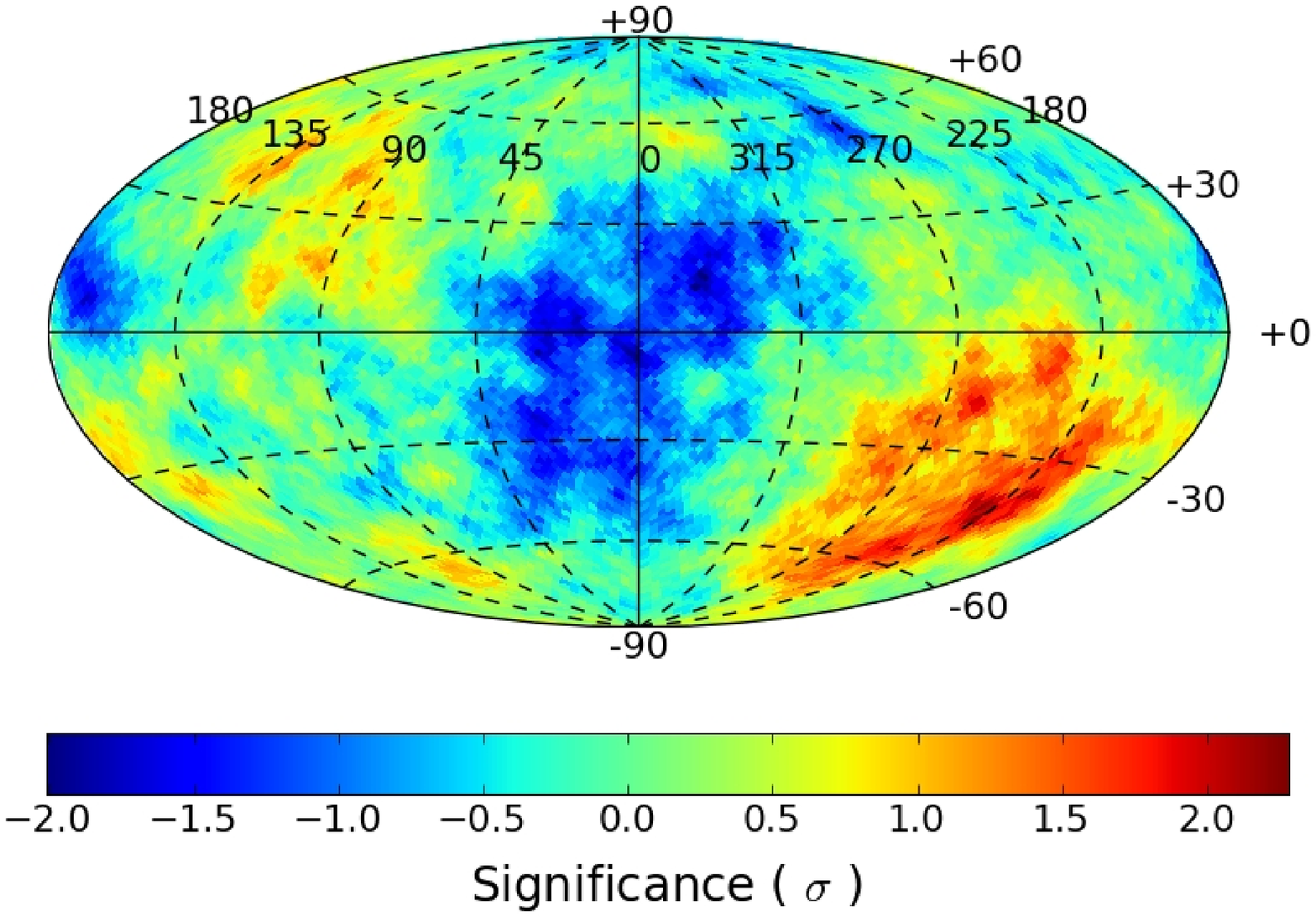}
\begin{center}
\includegraphics[width=0.87\columnwidth,keepaspectratio,clip]{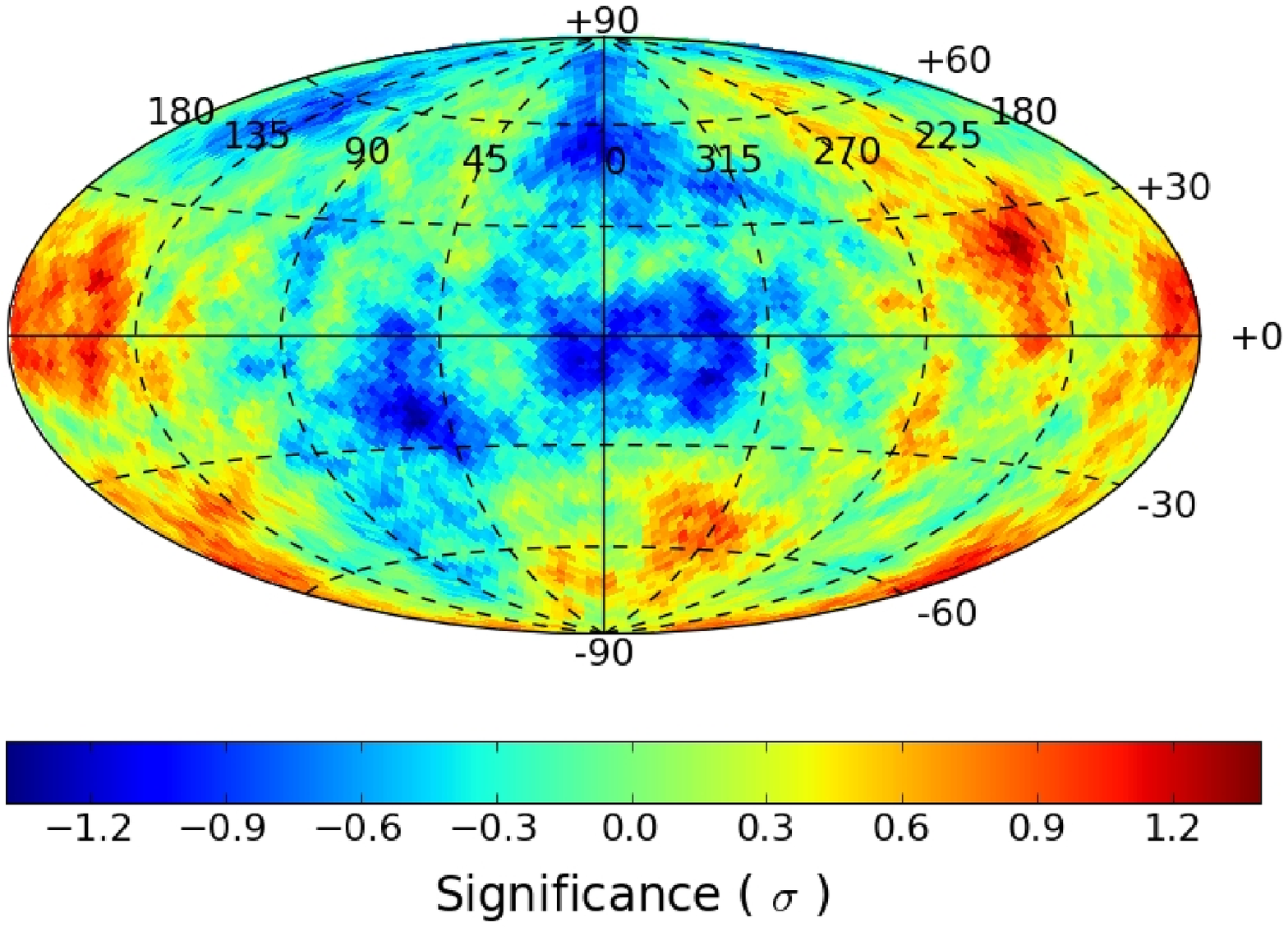}
\end{center}
\includegraphics[width=0.87\columnwidth,keepaspectratio,clip]{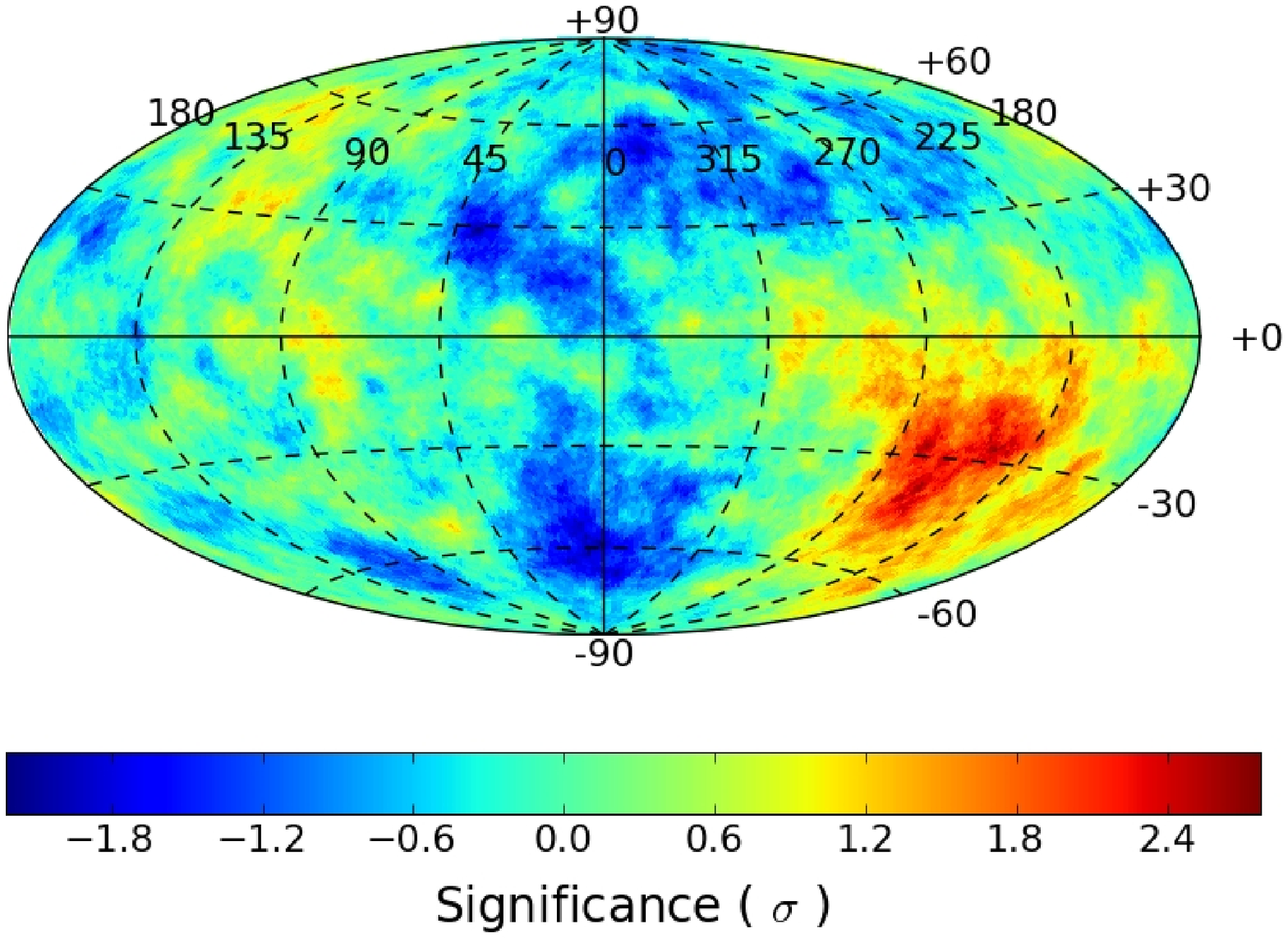}
\includegraphics[width=0.87\columnwidth,keepaspectratio,clip]{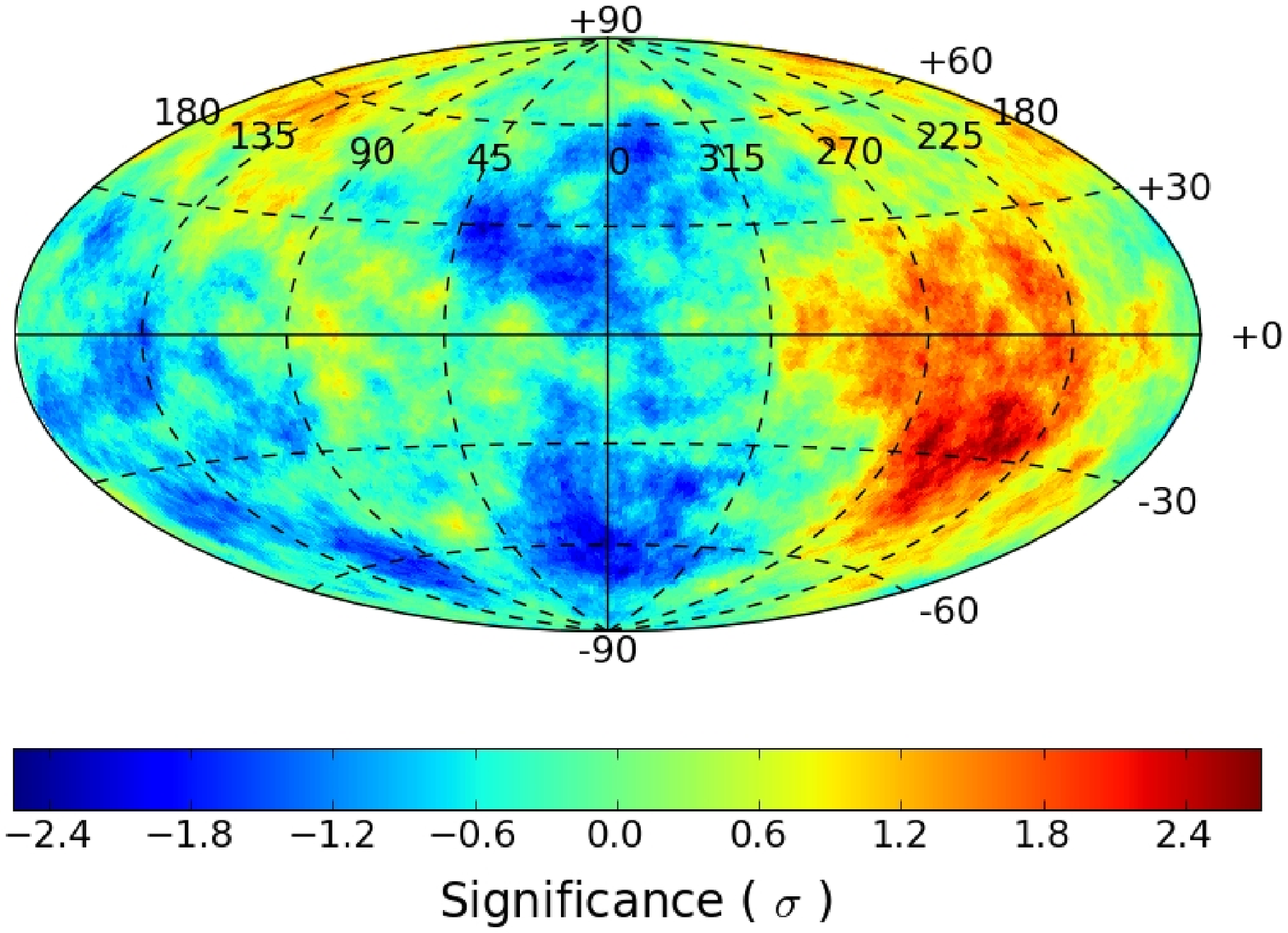}
\caption{Significance maps for E$>$60~GeV and for different integration radii. The first five plots were produced by the event shuffling technique and correspond to integration radii 10$^\circ$, 
30$^\circ$, 45$^\circ$, 60$^\circ$, and 90$^\circ$. The bottom two sky maps correspond to a 45$^\circ$ integration radius and were produced by the direct integration technique with $R_{allsky}(t)$ 
being given by the actual instantaneous CRE event rate (left) and the averaged-over-multiple-orbits CRE event rate (right).}
\label{figsigroi1}
\end{figure*}

\subsection{Bin to bin comparison}
Our no-anisotropy sky maps are initially produced consisting of independent bins. To use them 
for anisotropy searches we first integrate their bin contents over a range of angular scales 
ranging from $10^\circ$ to $90^\circ$.
One of the produced independent-bins no-anisotropy sky maps is shown in the first panel of Fig.~\ref{Fig_independent_60GeV}.
The corresponding actual map showing the actually detected events for the same energy range E$>$60~GeV
is shown in the second panel of the same figure. This sky map does not appear uniform because 
the sky was not observed with uniform exposure. The third panel shows the significance
sky map produced after comparing the other two sky maps.
The distribution of the values of this sky map with the Gaussian best-fit superimposed is shown in the top panel
of Fig.~\ref{fig3}. As expected, assuming the absence of any strong anisotropies, the best-fit function was statistically 
consistent with a Gaussian distribution of mean zero and unit variance. The same was also true for the 
E$>$120~GeV, E$>$240~GeV, E$>$480~GeV (not shown) distributions. 

Multiple pairs of signal and no-anisotropy sky maps were produced by integrating the independent-bin sky maps
(e.g. such as those shown in the first two panels of Fig.\ref{Fig_independent_60GeV}) over 
circular regions of radii ranging from 10$^\circ$ to 90$^\circ$. 
Similarly to the above, significance maps were constructed by
comparing the integrated signal and no-anisotropy sky maps. 
Figure \ref{figsigroi1} shows some of the significance sky maps obtained by comparing
the integrated no-anisotropy sky maps produced with the shuffling technique to the actual sky maps. 
The same figure also shows significance sky maps produced by the direct-integration technique and for a 45$^\circ$
integration radius.

\begin{figure}
\includegraphics[width=1\columnwidth,keepaspectratio,clip,trim=0 0 15 0]{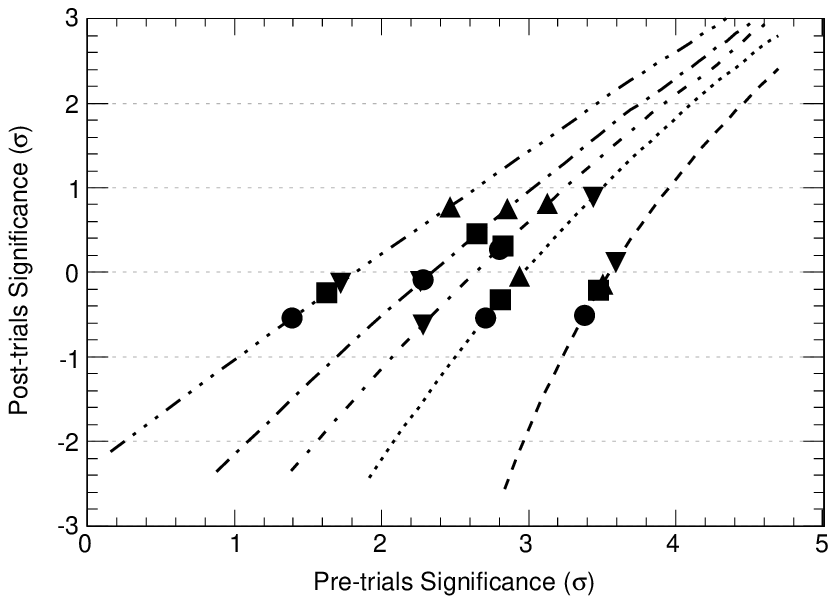}
\caption{Curves: correspondence between the pre-trials and significances for integrated maps of different integration radii;
from right to left 10$^\circ$, 30$^\circ$, 45$^\circ$, 60$^\circ$, and 90$^\circ$. Points: highest significances in each of the integrated sky maps (one point per minimum energies and integration radius): $\bullet$, E$>$60~GeV; $\blacksquare$, E$>$120~GeV; $\blacktriangle$, E$>$240~GeV; $\blacktriangledown$, E$>$480~GeV. The highest significance values (points) were all post trials insignificant.}
\label{SigsPrePost}
\end{figure}

The significances shown in these maps are pre-trials, i.e. they do not take into account the fact that we performed multiple trials while evaluating
them. Using the effective numbers of trials shown in Fig.~\ref{fig:Trials} and
Eq.~\ref{eq:p_post} we have calculated the post-trials significances
corresponding to the single most significant bin in each of the integrated maps
(one bin per integration radius and minimum energy), shown in Fig.~\ref{SigsPrePost}. Also shown in
the same figure is the correspondence between pre- and post-trials
significances for each integration radius. As can be seen, our most-significant
bins are all post-trials insignificant \cite{foot_6}.
It should be noted that there is one
more trial factor in this search corresponding to searching in multiple
integration radii and minimum energies. This factor is of the order of few and
because it is negligible compared to the trials factors of searching in all
directions, it was not included in the calculations. The inclusion of this trial
factor would just reinforce the null result of this search.

\begin{figure}
\includegraphics[width=1\columnwidth,keepaspectratio,clip,trim=0 0 15 0]{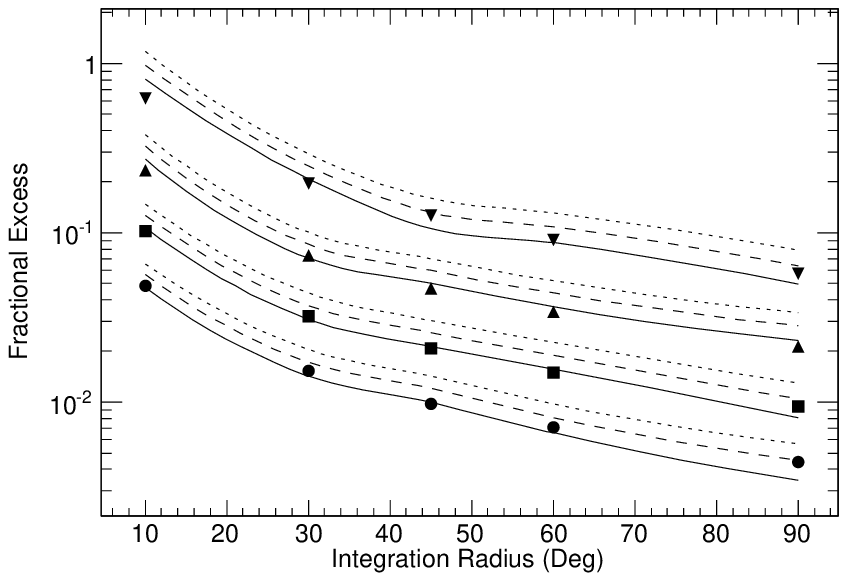}
\caption{Points: Fractional excess needed for the bin-to-bin comparison technique to detect an anisotropy with a post-trials significance 3$\sigma$ versus the integration radius and the minimum energy: $\bullet$, E$>$60~GeV; $\blacksquare$, E$>$120~GeV; $\blacktriangle$, E$>$240~GeV; $\blacktriangledown$, E$>$480~GeV. Curves: upper limits on the fractional excess. Each group of curves corresponds to a minimum energy: from bottom to top E$>$60~GeV, E$>$120~GeV, E$>$240~GeV, and E$>$480~GeV. Each line type corresponds to a different confidence level: solid, 1$\sigma$; dashed, 2$\sigma$; dotted 3$\sigma$.}
\label{fig:SensiUL}
\end{figure}

We have also performed the evaluation of our significance maps towards the direction of a few selected sources and regions. Such a search involves a considerably smaller number of trials since here only few specific directions are evaluated instead of the whole sky. We have searched for excesses towards the directions of the Vela ($l,b$ = 264$^\circ$, -3$^\circ$), Geminga (205$^\circ$, -1$^\circ$), and Monogem (201$^\circ$, 8$^\circ$) pulsars, towards the Virgo (300$^\circ$, 60$^\circ$) and Cygnus (75$^\circ$, 0$^\circ$) regions, and towards the Galactic and Galactic anticenter. To be conservative and to avoid accumulating too many trials, we searched for two different minimum energies, E$>$60~GeV and E$>$240~GeV, and for four different integration radii: 10$^\circ$, 30$^\circ$, 60$^\circ$, and 90$^\circ$. The evaluated sample did not show any significant anisotropies; the highest pre-trials significance value was from the direction of the Galactic anticenter: 3.03$\sigma$ for E$>$240~GeV and $10^\circ$ integration radius. The number of trials for this search is close to the total number of searches ($7\times2\times4=56$) and most likely lower since these searches are also correlated. Using the maximum number of trials (56), we find a best post-trials probability of $6.6\times10^{-2}$, or equivalently of 1.5$\sigma$, which is not significant. We also applied a stacking analysis in which we added the measurements from the three pulsars for the two minimum energies and integration radii (total four trials). This search also produced no significant results. 

Based on the absence of a detection, upper limits on anisotropies of angular scales
ranging from $10^\circ$ to $90^\circ$ from any direction in the sky were derived. The quantity constrained was the fractional excess, defined
as the number of excess anisotropy events from some circular region (spherical cap) in the sky over 
the number of events expected to be detected from the same region if the sky was perfectly isotropic. 
Figure~\ref{fig:SensiUL} shows the fractional excess needed to detect an anisotropy with a post-trials significance
of 3$\sigma$ versus the minimum energy and the integration radius. For larger energies or for smaller integration radii there are fewer events under an
integration region, therefore, a larger fractional excess is needed to produce a significant detection. In this figure,
also shown are the one side 1$\sigma$ (CL=84.1\%), 2$\sigma$ (CL=97.7\%), and 3$\sigma$ (CL=99.9\%) confidence level upper limits on the fractional excess.

\subsection{Spherical harmonic analysis}

\begin{figure*}
\includegraphics[width=1\columnwidth,keepaspectratio,clip,trim=0 0 30 0]{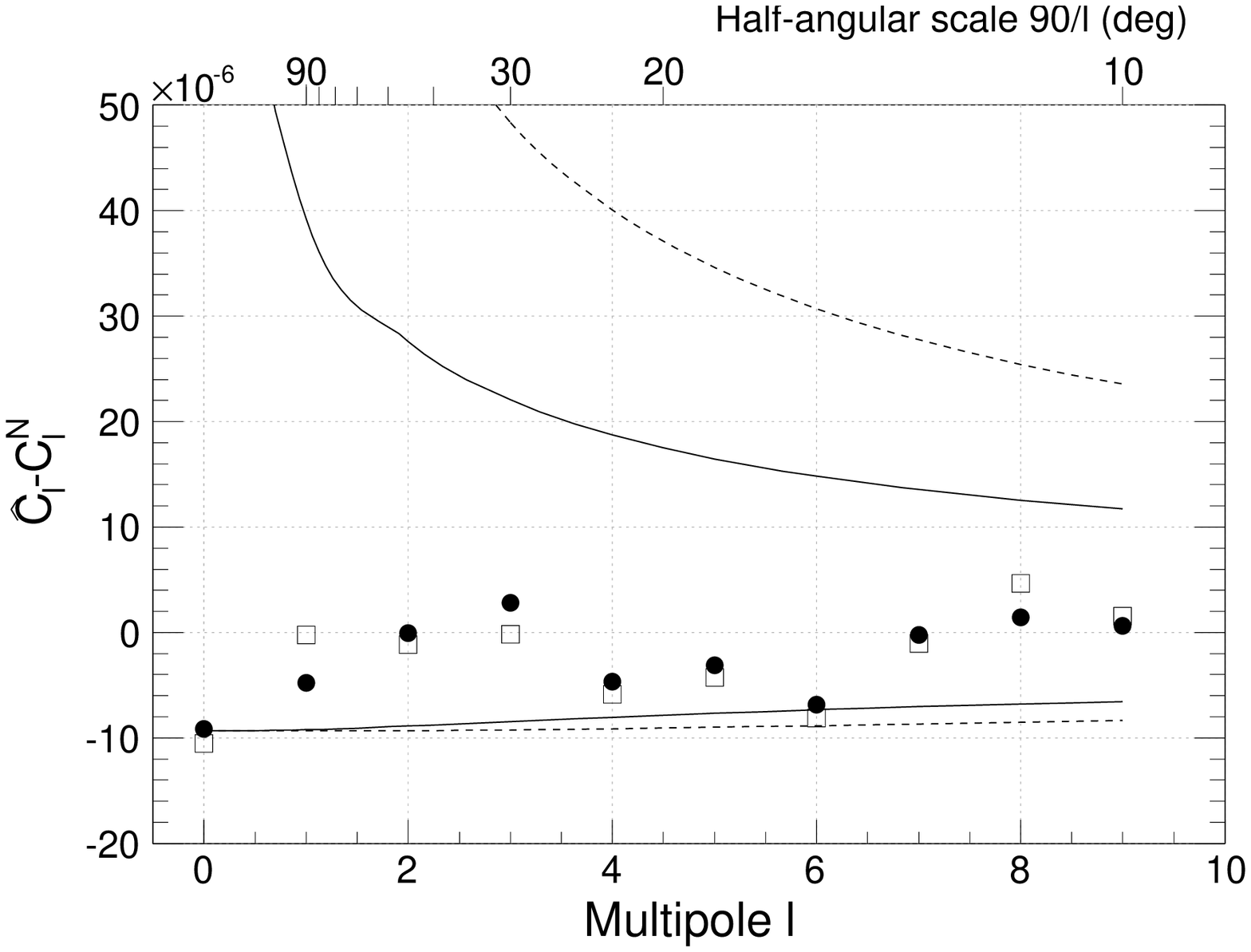}
\includegraphics[width=1\columnwidth,keepaspectratio,clip,trim=0 0 30 0]{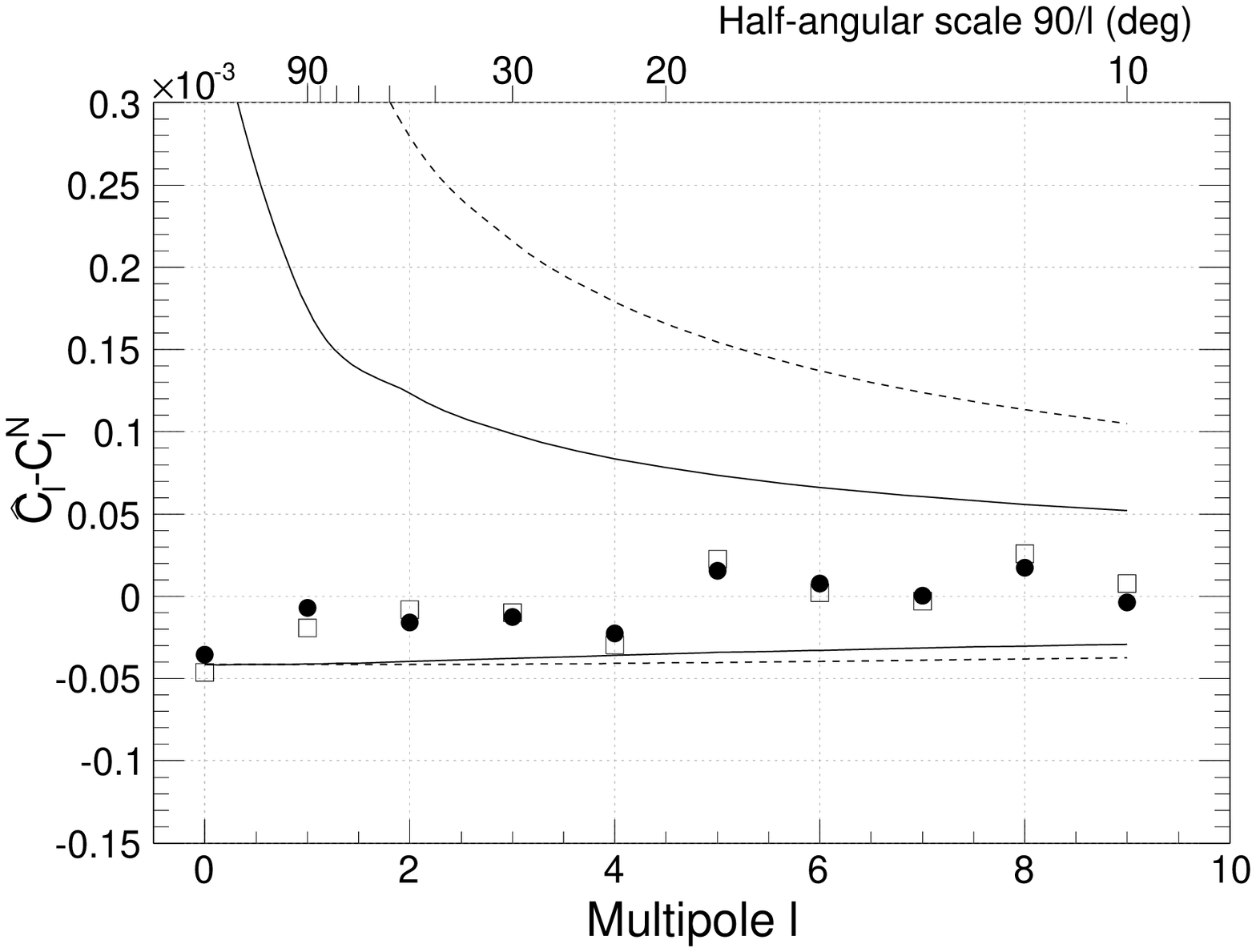}
\includegraphics[width=1\columnwidth,keepaspectratio,clip,trim=0 0 30 0]{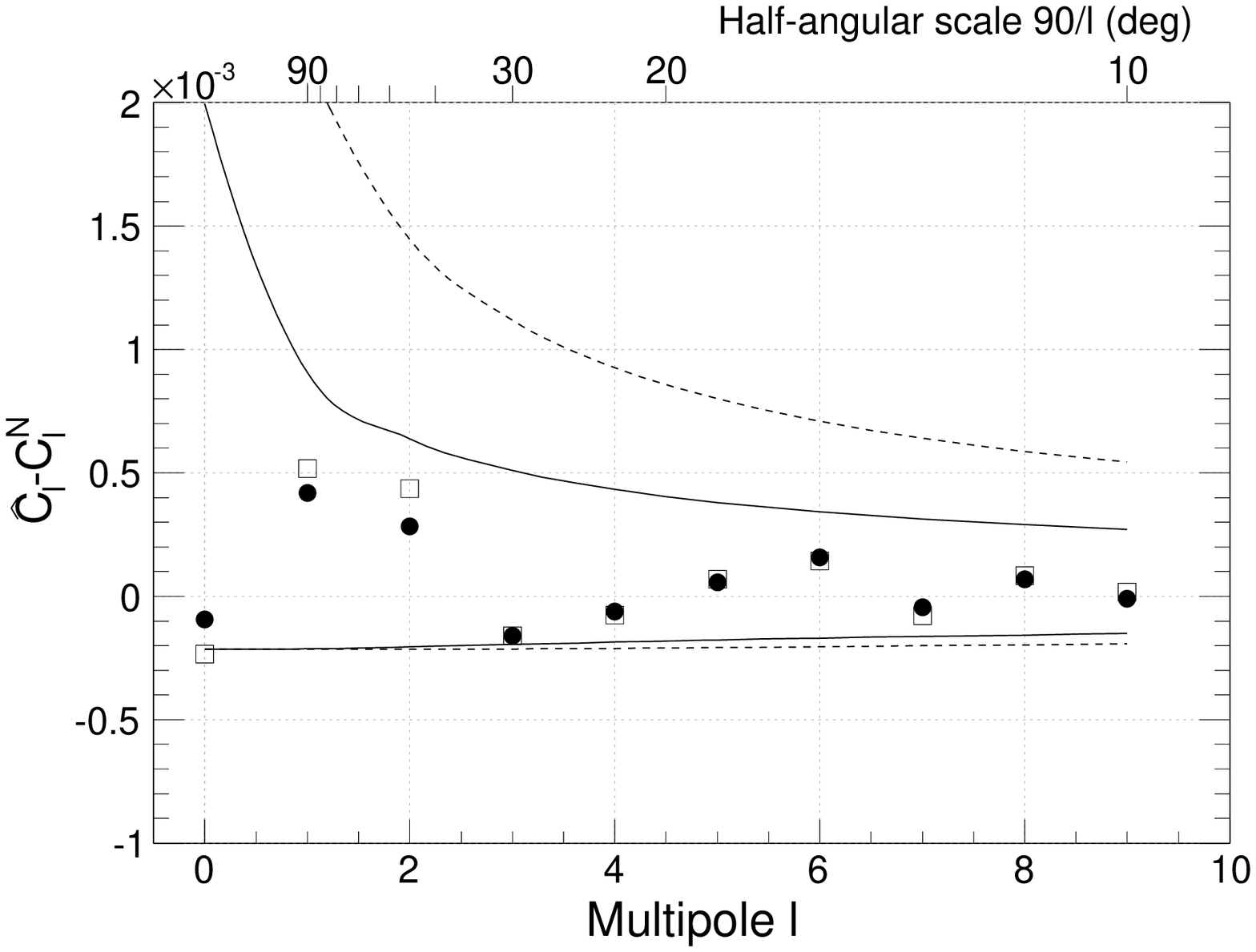}
\includegraphics[width=1\columnwidth,keepaspectratio,clip,trim=0 0 30 0]{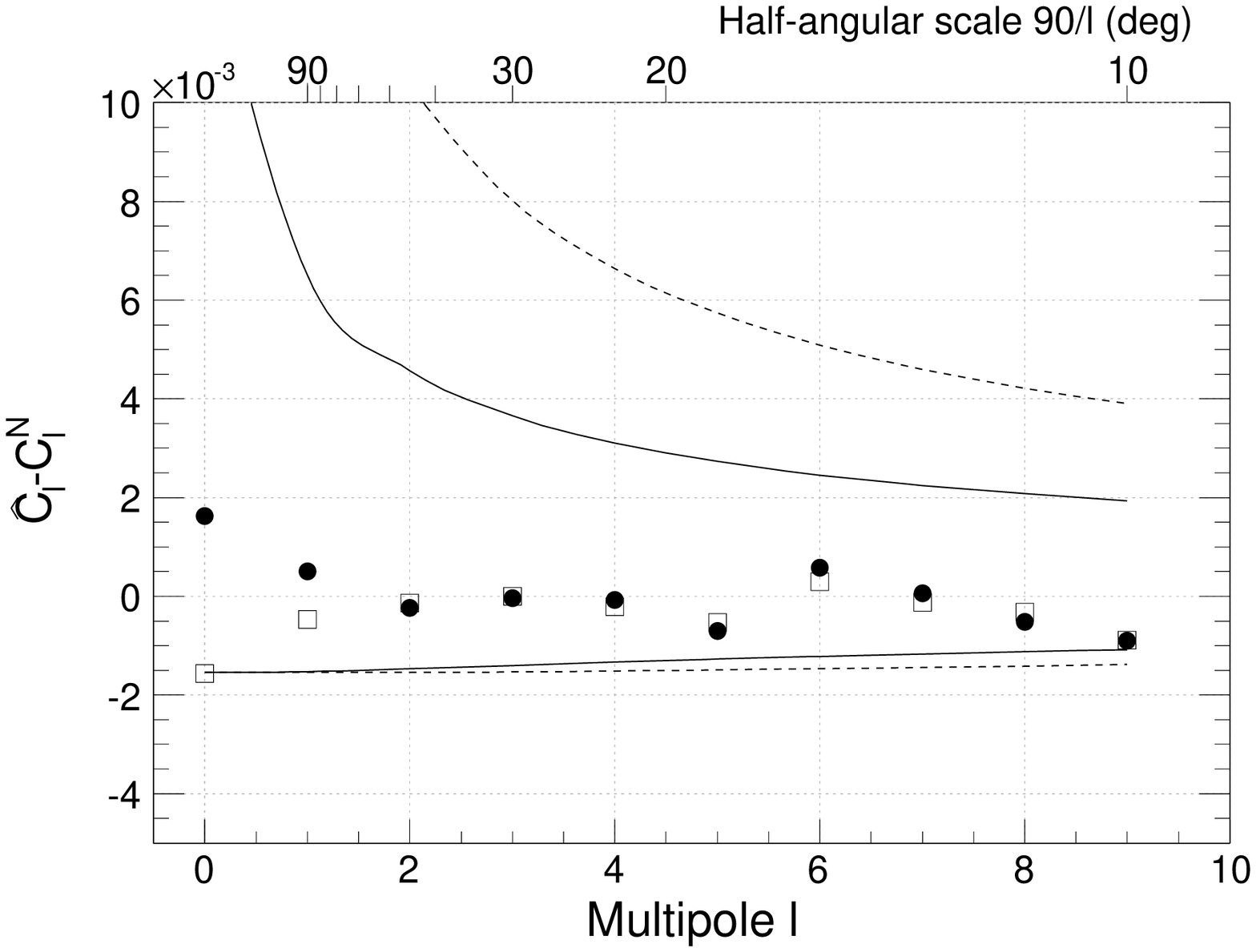}
\caption{Angular power spectra for different minimum energies;  top left, E$>$60~GeV; top right, E$>$120~GeV; bottom left, E$>$240~GeV; bottom right E$>$480~GeV. The points show the quantity $\hat{C}_{l}-C_{l}^{N}$ as produced by the event-shuffling (dots) and direct-integration (squares) techniques respectively. The two ranges show the 3$\sigma$ (solid) and 5$\sigma$ (dashed) interval of the probability distribution of the white-noise power spectrum $\hat{C}_{l}^{N}$. The data points of all spectra lie under the corresponding 3$\sigma$ ranges, showing that our measurements are consistent with the absence of any anisotropies.}
\label{PowerSpectra}
\end{figure*}

\begin{figure}
\includegraphics[width=1\columnwidth,keepaspectratio,clip,trim=0 0 20 0]{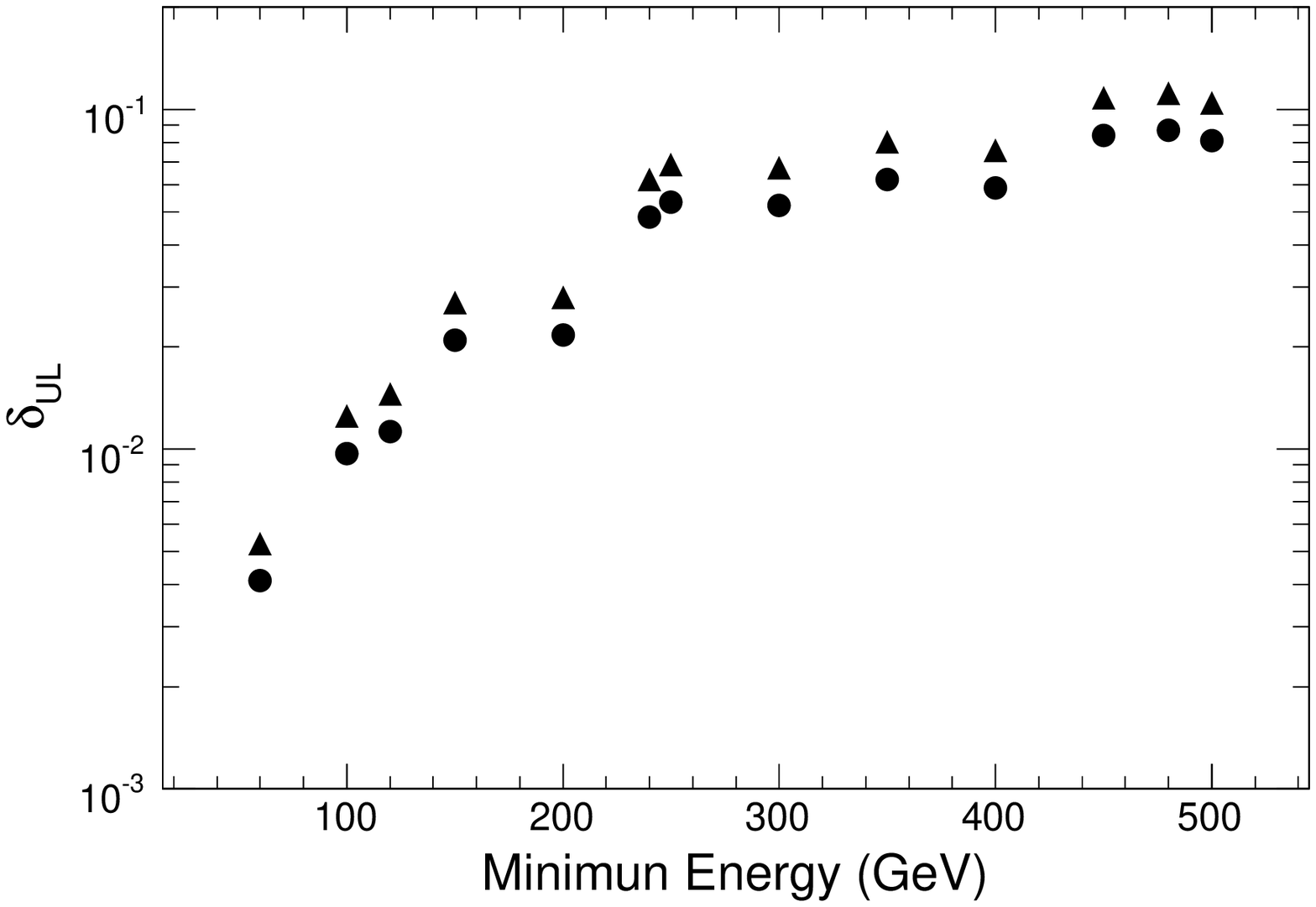}
\caption{Upper limits on the dipole anisotropy $\delta$ versus the minimum energy for different confidence levels; $\bullet$, 90 \% CL; $\blacktriangle$, 95\% CL.}
\label{fig:PowerSpectrumUL}
\end{figure}

The spherical harmonic analysis has been performed without removing the monopole term from the map.
Fig.~\ref{PowerSpectra} shows the angular power spectra for four different minimum energies, i.e. E$>$60~GeV, E$>$120~GeV, E$>$240~GeV and E$>$480~GeV. The markers in Fig. \ref{PowerSpectra} show the produced angular power spectra with the true value of the noise power spectrum subtracted, i.e. they are showing $\hat{C}_{l}-C^{N}_{l}$. The curves in the same figure show the range of the statistical fluctuations of the white-noise power spectrum ($\hat{C}^{N}_{L}$) for different integrated probabilities. All the data points (markers) lie inside the 3$\sigma$ range, showing that our measurements are consistent with an isotropic sky. Figure~\ref{fig:PowerSpectrumUL} shows the upper limits on the dipole anisotropy ($\delta$) as calculated using the $\hat{C_1}$ values of the event-shuffling technique as function of minimum energies from E$>$60~GeV to E$>$500~GeV 
\cite{foot_7}.

\section{Discussion}
\protect{\label{Sect_Disc}}

Cosmic-ray propagation in the Galaxy is often described assuming the diffusion approximation \cite{Ginzburg}. 
In this framework the dipole anisotropy due to the pure diffusion term is given by:

\begin{equation}
\delta = \dfrac{3 D}{c} \dfrac{|\vec{\nabla} N |}{N}
\label{eq:delta0}
\end{equation}
where $N$ is the density of particles and $D$ is the diffusion coefficient, that depends on the energy. 
Assuming a pure diffusive model, the spectrum of electrons and positrons from a point source can be calculated by solving the transport equation (see for example \cite{grasso}). In the case of a single source of age $t_i$ at the position $\vec{r}_i$, the contribution to the anisotropy is given by:

\begin{equation}
 \delta_i = \dfrac{3 D}{c} \dfrac{2 |\vec{r}_i|}{r^2_{diff}} 
 \label{eq:delta1}
\end{equation}
for $E<E_{max}$, where
$E_{max}$ is the maximum energy of the observed CREs originating from a given source due to its age and the rate of the energy loss ($E_{max}=1/(b_0~t_i)$, where $b_0 \simeq 1.4 \times 10^{-16} ~GeV^{-1}~s^{-1}$ ) \cite{foot_8}, and $r_{diff}$ is the diffusion distance that depends on the energy, the source age, and the diffusion coefficient~\cite{grasso}. 
The anisotropy of a single source evaluated from the above equation slightly decreases with the energy, and drops off for energies greater than $E=E_{max}$.
For $E \ll E_{max}$, the diffusion distance can be approximated as
$r_{diff}\simeq 2 \sqrt{D~t_i}$, and Eq.~\ref{eq:delta1} becomes:

\begin{equation}
 \delta_i = \dfrac{3 |\vec{r}_i|}{2 c t_i}.
\end{equation}

The total anisotropy due to a distribution of sources in the sky is then given by:

\begin{equation}
 \delta = \dfrac{\sum_i N_i~\delta_i~ \hat{r}_i \cdot \hat{n}_{max} }{\sum_i N_i}, 
\label{eq:delta3}
\end{equation}
where $\hat{n}_{max}$ is the direction of maximum intensity and $N_i$ is the spectrum for each individual source.

Recent results from the PAMELA~\cite{pamela} experiment show that the ratio of positrons to electrons plus positrons (the positron fraction) increases sharply in the energy range from 10~GeV to 100~GeV, in a way that appears to be inconsistent with secondary sources, i.e. with the interaction of cosmic-ray nuclei with
the interstellar gas. In addition, the \textit{Fermi}-LAT collaboration \cite{electronpaper, fullelectronpaper} has performed measurements of the $e^{-}e^{+}$ spectra up to 1~TeV with an unprecedented statistical accuracy and found an index harder than that measured in previous experiments. Nearby astrophysical sources, such as pulsars and supernova remnants, could provide a possible explanation for these features. Indeed, the electron/positron emission from a few nearby (up to a few hundreds of parsec distance) pulsars may give rise to observable anisotropies.

\begin{figure}[ht]
\includegraphics[width=1\columnwidth,keepaspectratio,clip,trim=0 0 20 0]{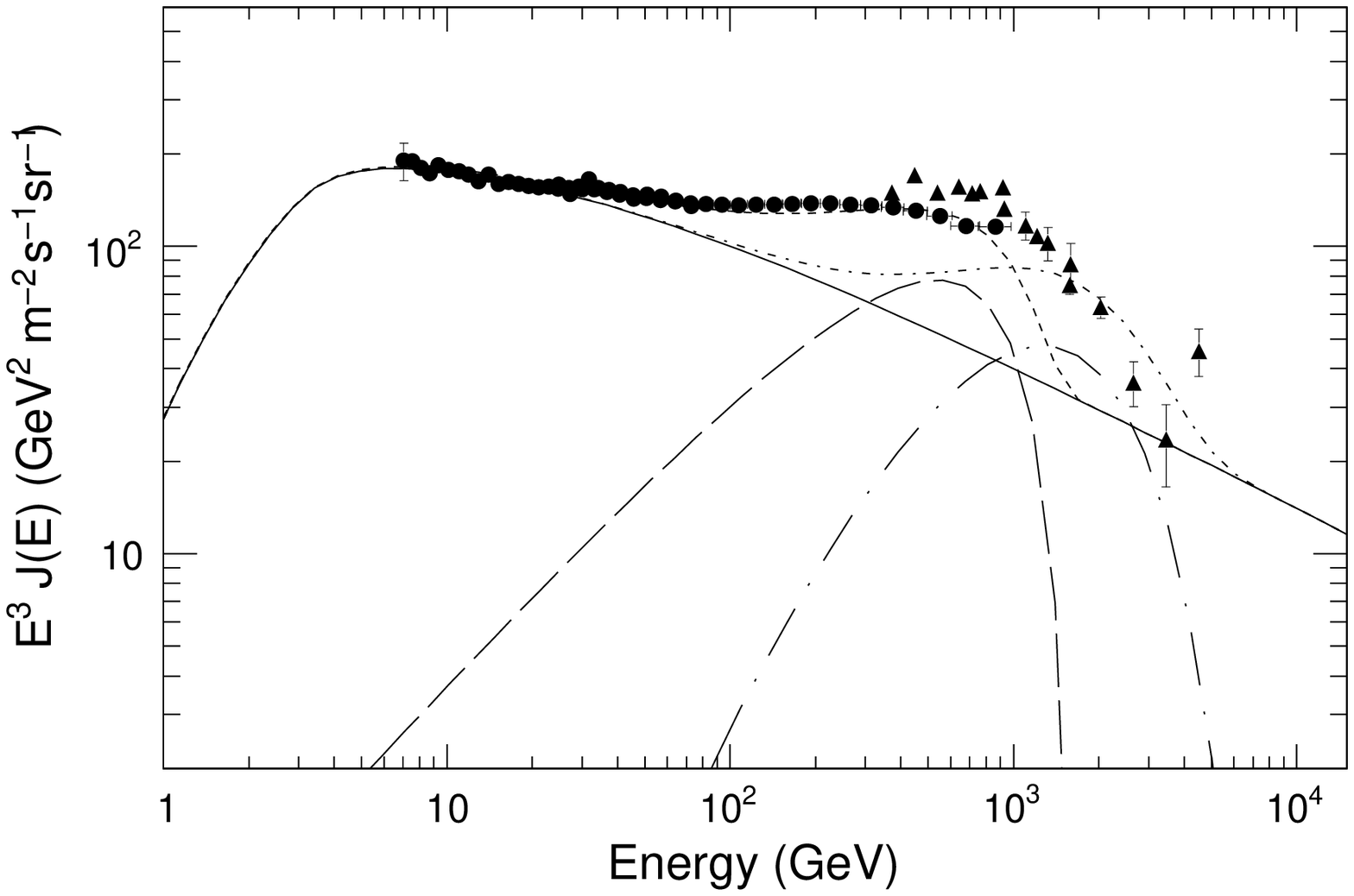}
\includegraphics[width=1\columnwidth,keepaspectratio,clip,trim=0 0 20 0]{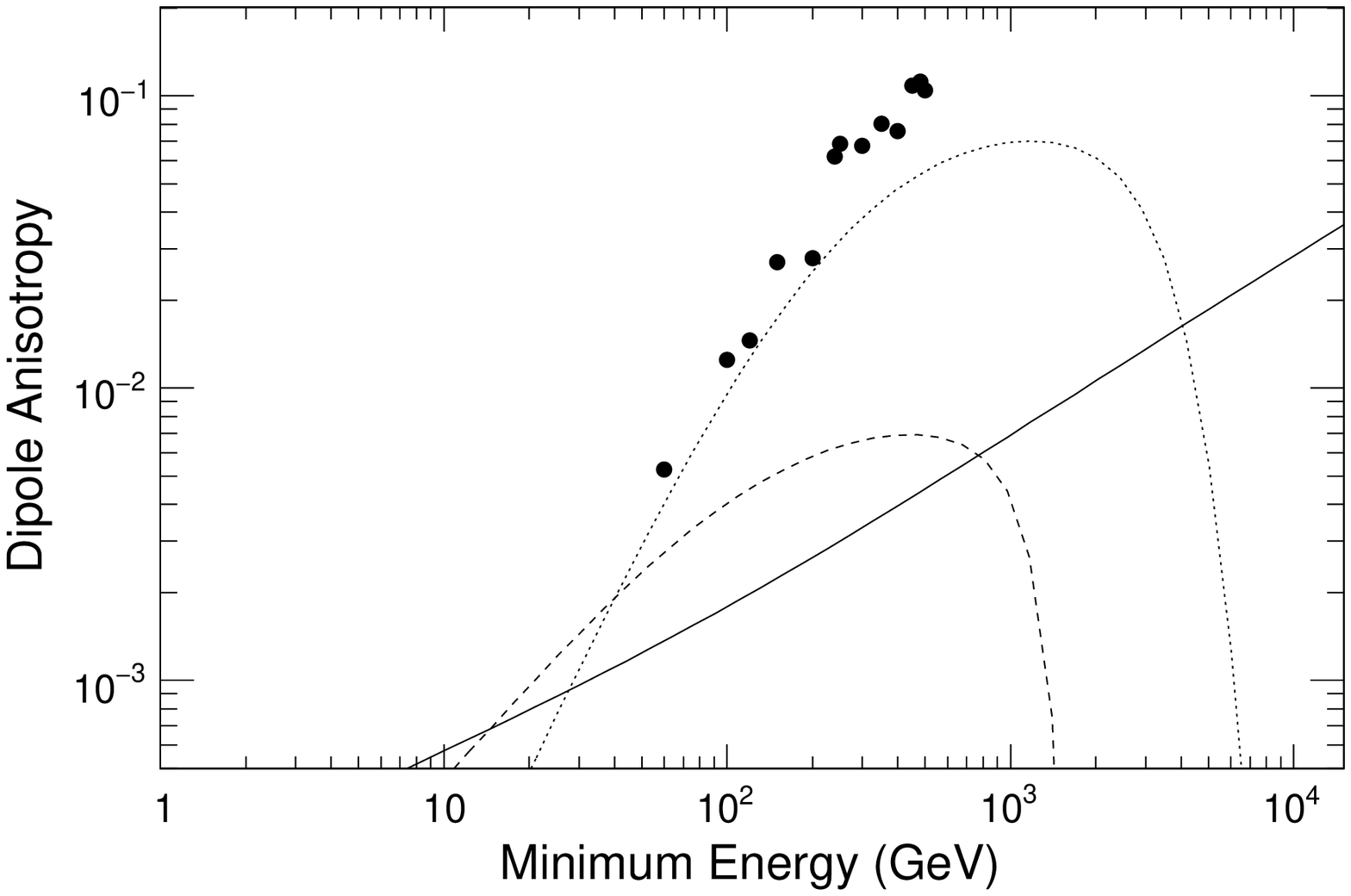}
\caption{Top panel: $e^+e^-$ spectrum evaluated with GALPROP and for single sources by means of Eq.~\ref{eq:sunspec}. Solid line: GALPROP spectrum; long dashed line: Monogem source; long-dot dashed line: Vela source; dashed line: GALPROP+Monogem; dot-dashed line: GALPROP+Vela; circles: \textit{Fermi}-LAT data~\cite{fullelectronpaper}; triangles: H.E.S.S. data \cite{hess08, hess09}. Bottom panel: Dipole anisotropy $\delta$ versus the minimum energy for GALPROP (solid line), Monogem source (dashed line), and Vela source (dotted line). The 95 \% CL from the data is also shown with circles. The solar
modulation was treated using the force-field approximation
with $\Phi$=550~MV~\cite{gleeson}.}
\label{fig:predictions}
\end{figure}

In order to evaluate the contribution of Galactic sources to the CRE anisotropy, we have performed a simulation with the GALPROP code \cite{galprop}, assuming a model that has been already used to interpret the CRE spectrum measured by the \textit{Fermi}-LAT \cite{fullelectronpaper}. In this model, the electron injection spectrum was assumed to be described by a broken power law with a spectral index of 1.6 (2.7) below (above) 4~GeV, and the diffusion coefficient was parameterized according to the usual power-law energy dependence $D(E)=D_0 (\frac{E}{E_0})^{0.33}$, where $D_0=5.8\times10^{28}$cm$^{2}$s$^{-1}$ and $E_0$~=~4~GeV. The diffusive reacceleration was characterized by an Alfven velocity $v_A$~=30~km$\,$s$^{-1}$ and the halo height was set to 4~kpc. 

The top panel of Fig.~\ref{fig:predictions} shows the GALPROP predictions for the CRE energy spectrum together with the \textit{Fermi}-LAT~\cite{fullelectronpaper} and H.E.S.S. data~\cite{hess08, hess09} (only statistical errors are shown). In the same panel, the fluxes expected from individual sources located in the Vela (290~pc distance and 1.1$\times 10^4$~yr age) and Monogem (290~pc distance and $1.1\times10^5$~yr age) positions are also shown. For the single sources, we have adopted an 
instantaneous injection spectrum of an electron source, i.e. a burst-like spectrum in which the duration of the emission is much shorter than the travel time from the source, 
described by a power law with index $\Gamma=1.7$ and with an exponential cut-off $E_{cut}$=1.1~TeV, i.e. $Q(E)=Q_0~E(GeV)^{-\Gamma}~exp(-E/E_{cut})$.
The spectrum of CREs at the solar system can be evaluated from the following equation~\cite{grasso}:

\begin{eqnarray}
& N(E, t_i, \vec{r}_i) = \dfrac{Q_0}{\pi^{3/2} r_{diff}^3}\left( 1-\dfrac{E}{E_{max}}\right)^{\Gamma-2} \left( \dfrac{E}{1GeV}\right) ^{-\Gamma} \nonumber \\
 & \exp\left(-\dfrac{E}{(1-\frac{E}{E_{max}})E_{cut}}\right)~\exp\left( -\dfrac{r_i^2}{r_{diff}^2}\right).
  \label{eq:sunspec}
\end{eqnarray}
For both sources, the value of the normalization constant $Q_0$ has been chosen to obtain a total flux not higher than that measured by the \textit{Fermi}-LAT and H.E.S.S. (top panel of Fig.~\ref{fig:predictions}). 

The bottom panel of Fig.~\ref{fig:predictions} shows the dipole anisotropy as a function of minimum energy, calculated using the $e^+ e^-$ spectrum evaluated with GALPROP by means of Eq.~\ref{eq:delta0}. In the same panel, the dipole anisotropies expected from the Monogem and Vela sources are also shown. For each source, the anisotropy has been evaluated by means of Eq.~\ref{eq:delta3}, where we have assumed that the contributions to the anisotropy from all remaining sources are negligible, that is, $\hat{n}_{max}=\hat{r}_i$ where $\hat{r}_i$ is the direction of the source under investigation, and $\delta_j=0$ for $j \neq i$. It is worth to point out that in the denominator of Eq.~\ref{eq:delta3} the Monogem (Vela) source is added to the total CRE flux evaluated with GALPROP. Moreover, the dipole anisotropy above a given energy is evaluated as the ratio between the integral in energy of the numerator and the integral in energy of the denominator of Eq.s~\ref{eq:delta0} and \ref{eq:delta3}. This comes from the definition of the degree of the anisotropy shown in Eq.~\ref{eq:dip}, where the intensities are integrated above a given energy.

According to the above predictions, the level of anisotropy expected for Vela-like and Monogem-like sources (i.e. sources with similar distances and ages) is not excluded by the results shown in Fig.s~\ref{fig:SensiUL} and \ref{fig:PowerSpectrumUL}. However, it is worth pointing out that the model results are affected by large uncertainties related to the choice of the free parameters (i.e. $Q_0$, $E_{cut}$, and $\Gamma$). 

The positron excess detected by PAMELA can be ascribed not only to astrophysical sources such as pulsars, but also to the annihilation or decay of Galactic dark matter (see e.g. \cite{grasso}). Interestingly, as pointed out in the early analyses (see for example \cite{Hooper:2008kg, Profumo:2008ms}), any anisotropy in the arrival directions of CREs detected by the LAT is a powerful tool to discriminate between a dark matter origin and an astrophysical one. In particular, since Galactic dark matter is denser towards the direction of the Galactic center, the generic expectation in the dark matter annihilation or decay scenario is a dipole with an excess towards the center of the Galaxy and a deficit towards the anti-center. Luckily, as pointed out in \cite{Hooper:2008kg}, both the Monogem and the Geminga pulsars, likely some of the most significant CRE pulsar sources, even after the discovery of several radio-quiet gamma-ray pulsars by the LAT \cite{Gendelev:2010fd}, are both roughly placed opposite to the direction of the Galactic Center, making a search for anisotropy an effective distinguishing diagnostic.

\begin{figure}[ht]
\includegraphics[width=1\columnwidth,keepaspectratio,clip,trim=0 0 20 0]{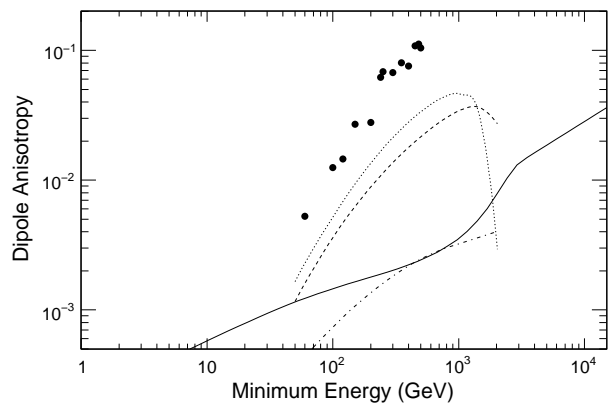}
\caption{Dipole anisotropy $\delta$ versus the minimum energy for some DM scenarios. Solid line: DM distributed in the Milky Way Halo; dashed and dotted lines: two dark matter benchmark models taken from \cite{Regis:2009qt};  dot-dashed line: DM from the population of Galactic substructures \cite{Cernuda:2009kk} (see text). The 95 \% CL upper limits on the dipole anisotropy from the data are also shown with circles.}
\label{fig:Regis}
\end{figure}

The expected level of dipole anisotropy produced by dark matter annihilating in the Milky Way halo, calculated by tuning the annihilation rate to match the positron fraction measured by the PAMELA satellite, is comparable or more likely smaller than the degree of anisotropy expected by astrophysical Galactic sources as modeled in GALPROP (see the solid line in the bottom panel of Fig.~\ref{fig:predictions}). 
We verified this with an explicit calculation with GALPROP, slightly modified to include the injection of CREs from DM annihilation while using the same propagation setup employed to derive the anisotropy from nearby pulsars.
The GALPROP results from the conventional astrophysical Galactic sources and from a scenario with DM distributed
according to a Navarro, Frenk and White (NFW) profile, with a 3~TeV mass candidate that annihilates into $\tau^+ \tau^-$ with a cross section of $\langle \sigma v \rangle = 5 \times 10^{-23}$ cm$^3$ s$^{-1}$, a local DM density of 0.43~GeV~cm$^{-3}$ and a 20~kpc of core radius have been added. With this DM model, the measured overall CRE flux by the LAT and the charge ratio measured by PAMELA are reproduced. The solid line in Fig.~\ref{fig:Regis} shows the total expected anisotropy level, which is similar to that predicted when only the astrophysical sources are considered in GALPROP (see the solid line in the bottom panel of Fig.~\ref{fig:predictions}). 

A caveat however exists to the statement above, due to the possibility that most of the high-energy positrons detected by PAMELA are produced by dark matter annihilations in a nearby dark matter clump. The halo of the Milky Way, in the context of the cold dark matter paradigm, is in fact thought to host a myriad of hierarchical smaller sub-halos and sub-sub-halos, potentially contributing significantly to the dark matter annihilation signal, as envisioned in \cite{Hooper:2008kv,Bringmann:2009ip,Kuhlen:2009is,Brun:2009aj}. In the analysis of \cite{Brun:2009aj}, it was shown that (a) compared to N-body simulation results \cite{Diemand:2008in}, the likelihood of a nearby and luminous clump that could explain the PAMELA excess is very remote (to the level of less than 0.01\%) for ordinary pair-annihilation cross-sections, and (b) when assuming large annihilation cross-sections, the predicted associated gamma-ray flux from dark matter annihilation would in most cases exceed the point-source sensitivity of the LAT. In other words and for the second point, if a clump is responsible for most of the locally measured positrons, it would have very likely already been observed it shining in gamma rays.

To illustrate the anisotropy from single nearby dark matter clumps, we take two benchmark
models from ~\cite{Regis:2009qt} that give good fits to the PAMELA and \textit{Fermi} data. In these models, the clumps are moving
with a speed of 300~km~s$^{-1}$ perpendicular to the Galactic plane, and the dark matter particle has a
mass of 5 (3) TeV annihilating into $\tau$ leptons. Figure~\ref{fig:Regis} shows the anisotropy induced by
such a point source departing at 1.54 kpc (approaching at 1.43 kpc) as a dashed (dotted) line.
The anisotropy is typically mainly sensitive to the dark matter clumps' distance, but the different
anisotropies for these two clumps (at almost the same distance) are mainly due to an
approaching compared to a departing clump.
It should be noted that the diffusion models considered in \cite{Regis:2009qt} are derived assuming plane diffusion, with the diffusion coefficient $D(E)=D_0 \times (E/E_0)^{0.6}$, $E_0=4$~GeV and $v_A=0$~km$\,$s$^{-1}$. The two models in Fig.~\ref{fig:Regis}, assume halo heights and corresponding values of the diffusion constant $D_0$ of $z=4$~kpc, $D_0=2.5 \times 10^{28}$ cm$^2$s$^{-1}$ (dashed curve) and $z=10$ kpc, $D_0= 4.6 \times 10^{28}$ cm$^2$s$^{-1}$ (dotted curve). The local electron spectrum from astrophysical sources is assumed to have a soft power law index of $3.24$, which under-predicts the \textit{Fermi} CRE spectrum measurements at the highest energies.

For comparison, we also show in Fig.~\ref{fig:Regis} the expected anisotropy from a population of
Galactic substructures, as calculated in ~\cite{Cernuda:2009kk} (dot-dashed line). In this calculation the
distribution of dark matter clumps in the halo was assumed to follow a NFW profile, as in ~\cite{Cumberbatch:2006tq}, and the dark matter particle has a mass of 3.6~TeV and annihilates into $\tau$ leptons. It should be noted that the dark-matter-induced anisotropies predicted here are only valid within their given
set-up; additional significant CRE-source contributions, a different astrophysical background, or
a modified diffusion model, would modify the expected dipole-anisotropy signal.
We conclude that, even in optimistic scenarios of a bright and reasonable local dark matter
clump, the anisotropy signal is expected to be below the experimental
limits reported here.

The CG effect predicts a dipole amplitude of $(\Gamma+2)\frac{v}{c}$ for CRs with a power law spectrum $\sim E^{-\Gamma}$, and for an observer (Earth) moving with speed $v$ with respect to the local CR plasma. Two relative motions -- the motion of the Sun with respect to the CR plasma, and the motion of the Earth with respect to the Sun -- can create anisotropies through the CG effect. Previous studies of CR anisotropies failed to detect a CG effect due to the motion of the Sun, showing that the CR plasma co-rotates with the local stars \cite{tibet06,superK}. Therefore, in this study we only expect to detect a CG effect due to the motion of the Earth around the Sun. The expected amplitude of this effect (for an orbital speed of v=29.8~km~s$^{-1}$) is $\sim5\times10^{-4}$, considerably smaller than the sensitivity of this search. 
A dedicated analysis fully investigating the CG effect as function of both solar and sidereal time may be more sensitive in detecting it.

Contamination of the CRE sample with other species (protons) can introduce some systematic uncertainties in the measurement.
Ground experiments have detected anisotropies for protons of energies above 10~TeV at the $10^{-3}$ level. Since these fluctuations are expected to increase with energy at a rate of about $E^{0.1}$ \cite{ptuskin}, they are expected at a level of $10^{-4}-10^{-3}$ for the current observed energy band. We should expect fluctuations of the order of $10^{-4}$ or less due to proton contamination in our CRE selection, less than the sensitivity of this search. Finally, it should be noted that the results presented here are valid to the degree that the HMF did not smear away any actual Galactic anisotropies. It could be the case that the detected isotropy is the result of smearing by the HMF, instead of a truly isotropic Galactic CRE distribution.

\section{Conclusion}
More than 1.6 $\times$ $10^6$ primary cosmic-ray electrons/positrons with energies above 60~GeV have been observed by the LAT instrument during its first year of operation. An all-sky study without any a priori assumptions on the energy, direction, and angular size has been performed to search for possible anisotropies in the incoming directions of these events. The search was performed using two independent and complementary techniques, both providing a null result. The upper limits on a fractional anisotropic excess ranged from a fraction of a percent to roughly one, for the range of minimum energies and angular scales considered. A detailed study of the dipole anisotropy has been also performed, and upper limits ranging from $\sim0.5\%$ to $\sim 10\%$, depending on the energy, have been set.  Our upper limits on the dipole anisotropy were compared with the predicted anisotropies from individual nearby pulsars and from dark matter annihilations. In all cases, our upper limits lie roughly above the predicted anisotropies. It should be noted that the calculations of the predicted anisotropies involve a large number of free parameters. Therefore, our upper limits can be used to constrain the parameter space of these models excluding any combinations that are in conflict with our null result. 

\section*{Acknowledgements}
The \textit{Fermi} LAT Collaboration acknowledges generous ongoing support
from a number of agencies and institutes that have supported both the
development and the operation of the LAT as well as scientific data analysis.
These include the National Aeronautics and Space Administration and the
Department of Energy in the United States, the Commissariat \`a l'Energie Atomique
and the Centre National de la Recherche Scientifique / Institut National de Physique
Nucl\'eaire et de Physique des Particules in France, the Agenzia Spaziale Italiana
and the Istituto Nazionale di Fisica Nucleare in Italy, the Ministry of Education,
Culture, Sports, Science and Technology (MEXT), High Energy Accelerator Research
Organization (KEK) and Japan Aerospace Exploration Agency (JAXA) in Japan, and
the K.~A.~Wallenberg Foundation, the Swedish Research Council and the
Swedish National Space Board in Sweden.

Additional support for science analysis during the operations phase is gratefully
acknowledged from the Istituto Nazionale di Astrofisica in Italy and the Centre National d'\'Etudes Spatiales in France.


\begin{thebibliography}{99}
\bibitem{foot_1} The electron diffusion distance has been evaluated using the diffusion coefficient and the energy loss rate as defined in Sect.~\ref{Sect_Disc}.
\bibitem{foot_2} $\theta$ and $\phi$
are the off-axis and azimuth angles in instrument coordinates.
\bibitem{foot_3} Equal to the number of CRE events detected from the whole sky in a multiple-orbits-long time interval, i.e. at least several hours long, divided by the livetime of that interval.
\bibitem{foot_4} A positive sign of the significance is chosen in case of excess, and a negative sign is chosen in case of deficit.
\bibitem{foot_5}Consider an outcome $X$ that occurs with probability $P_{pre}(X)$ in a single trial. The probability
of the outcome not occurring in a single trial is $1-P_{pre}(X)$, in two independent trials $\left(1-P_{pre}(X)\right)^2$,
and in $T$ independent trials $\left(1-P_{pre}(X)\right)^T$. Consequently, the probability of 
this outcome occurring at least once in $T$ uncorrelated trials is $P_{post}(X,n>=1)=1-(1-P_{pre}(X))^T$. For correlated
trials $T$ is replaced by its effective number $T_{eff}$ and we obtain Eq.\ref{eq:p_post}. Solving for $T_{eff}$ leads to
Eq.\ref{eq:teff}.
\bibitem{foot_6} It should be noted that a significance S corresponds 
to a probability equal to the integral from S to infinity under a standard normal distribution. 
It could be the case that a positive post-trials significance, which corresponds to 
a pre-trials significance smaller than 0.5, be associated with a larger than 0.5 post-trials
probability, which corresponds to a negative post-trials significance.
\bibitem{foot_7} Despite the fact that the results up to now have been reported for the subsets of the data with minimum energies E$>$60~GeV, E$>$120~GeV, E$>$240~GeV, E$>$480~GeV, all the plots shown below will also include upper limits on the dipole anisotropy produced with minimum energies that are integer multiples of 50~GeV starting from 100~GeV up to 500~GeV.
\bibitem{foot_8} The $b_0$ value has been chosen to be consistent with that in GALPROP at the Sun position ~\cite{grasso}.
\end{thebibliography}
\end{document}